\documentclass[11pt,a4paper]{article}
\usepackage{jheppub}

\usepackage{amsmath,amssymb,amsfonts,bm,amscd}
\usepackage{mathtools}
\usepackage{graphicx}
\usepackage{multirow}
\usepackage{verbatim}
\usepackage{appendix}
\usepackage{fancybox}
\usepackage{array}

\newcommand{\bea}{\begin{eqnarray}}
\newcommand{\eea}{\end{eqnarray}}
\newcommand{\be}{\begin{equation}}
\newcommand{\ee}{\end{equation}}

\newcommand{\beq}{\begin{equation}}
\newcommand{\eeq}{\end{equation}}

\newcommand{\Z}{{\mathbb Z}}
\newcommand{\R}{{\mathbb R}}
\newcommand{\C}{{\mathbb C}}

\def\Tr{{\rm Tr \,}}

\def\G{\Gamma}
\def\e{\epsilon}

\def\tilde{\widetilde}
\def\hat{\widehat}
\def\bar{\overline}

\def\CF{{\mathcal F}}

\def\CI{{\mathcal I}}

\def\CL{{\mathcal L}}
\def\CM{{\mathcal M}}
\def\CN{{\mathcal N}}
\def\CO{{\mathcal O}}

\def\CU{{\mathcal U}}

\newcommand{\cp}{{\mathbb{C}}{\mathbf{P}}}

\renewcommand{\bar}{\overline}
\renewcommand{\hat}{\widehat}

\def\Tr{{\mathrm{Tr}}}

\title{``Lagrangian Disks'' in M-theory}

\author[a,b,c]{Sebasti\'an Franco,} 
\author[d]{Sergei Gukov,} 
\author[e,f,g]{Sangmin Lee,}
\author[h]{Rak-Kyeong Seong,}
\author[i]{James Sparks}

\affiliation[a]{Physics Department, The City College of the CUNY, 160 Convent Avenue, New York, NY 10031, USA} 
\affiliation[b]{Physics Program and $^c$Initiative for the Theoretical Sciences \\
The Graduate School and University Center, The City University of New York  \\
365 Fifth Avenue, New York NY 10016, USA}
\affiliation[d]{Walter Burke Institute for Theoretical Physics, California Institute of Technology, Pasadena, CA 91125, USA}
\affiliation[e]{Center for Theoretical Physics, Seoul National University, Seoul 08826, Korea}
\affiliation[f]{Department of Physics and Astronomy, Seoul National University, Seoul 08826, Korea}
\affiliation[g]{College of Liberal Studies, Seoul National University, Seoul 08826, Korea}
\affiliation[h]{Yau Mathematical Sciences Center, Tsinghua University, 100084 Beijing, China}
\affiliation[i]{Mathematical Institute, University of Oxford, Woodstock Road, Oxford, OX2 6GG, UK}

\abstract{While the study of bordered (pseudo-)holomorphic curves with boundary on Lagrangian submanifolds has a long history, a similar problem that involves (special) Lagrangian submanifolds with boundary on complex surfaces appears to be largely overlooked in both physics and math literature. We relate this problem to geometry of coassociative submanifolds in $G_2$ holonomy spaces and to $Spin(7)$ metrics on 8-manifolds with $T^2$ fibrations. As an application to physics, we propose a large class of brane models in type IIA string theory that generalize brane brick models on the one hand and 2d theories $T[M_4]$ on the other.
}

\preprint{
\begin{flushright}
{}
\end{flushright}
}

\begin{document}
\maketitle



\section{Introduction and summary}

A familiar setup, both in physics and in math,
involves a (special) Lagrangian submanifold $L$ in a Calabi-Yau 3-fold $X$
together with bordered pseudoholomorphic curve $\Sigma$ with boundary in $L$.
Exploring all possible choices of $\Sigma$ with fixed $X$ and $L$ is the main goal
of ``open Gromov-Witten theory'' and plays an important role in counting disk instantons \cite{MR2567952} which, in turn, are paramount for stability of string vacua.

\begin{figure}[ht]
	\centering
	\includegraphics[width=2.7in]{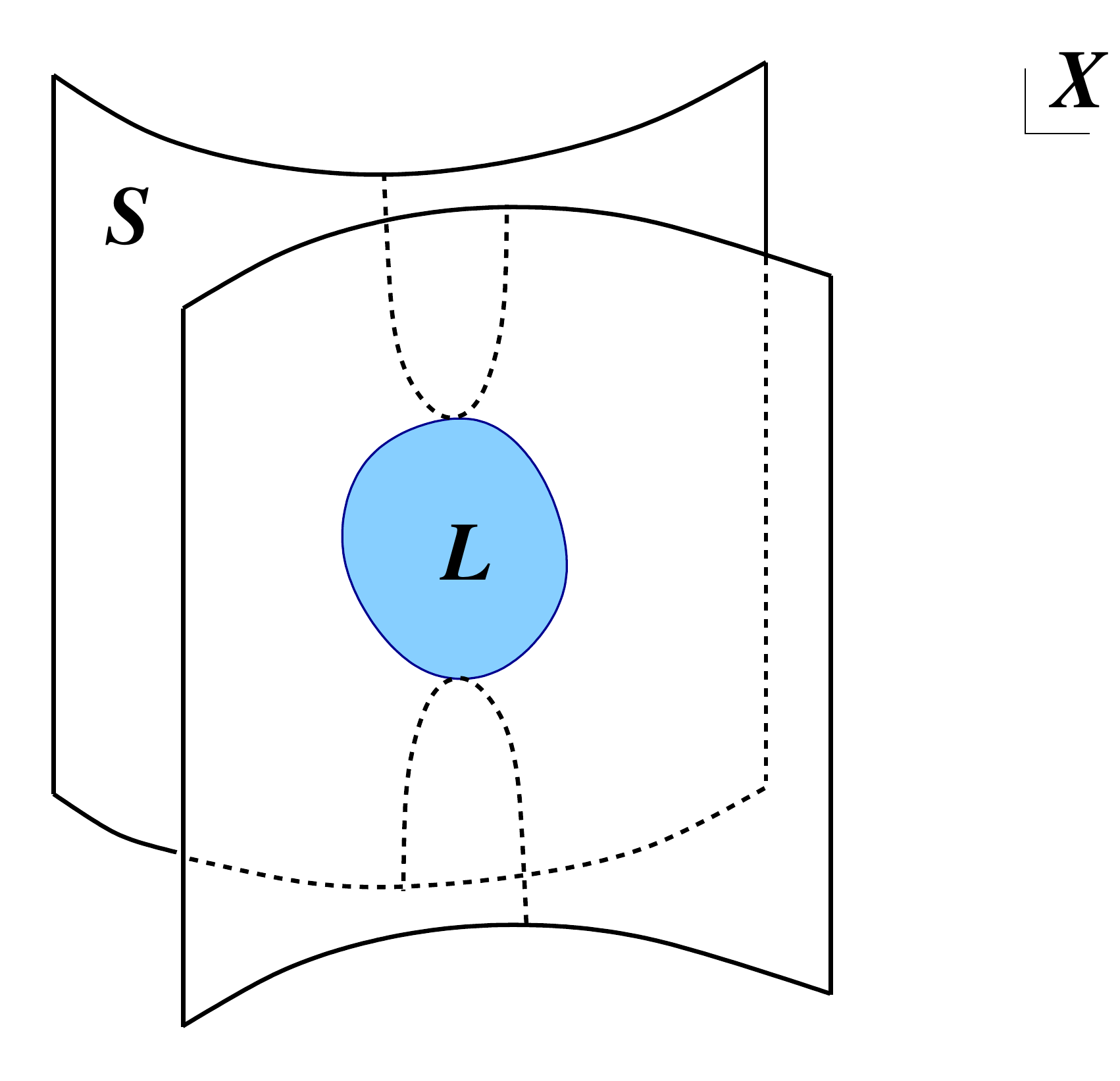}
	\caption{Our geometric setup involves a holomorphic surface $S$ in a Calabi-Yau 3-fold $X$,
		and a special Lagrangian submanifold $L \subset X$ with boundary $\Sigma := \partial L \subset S$.}
	\label{fig:XSL}
\end{figure}

In this paper we consider a much less explored version of this setup, which involves a similar triple, illustrated in Figure~\ref{fig:XSL}:
\be
\begin{array}{ll}
X: & \text{Calabi-Yau 3-fold} \\
S: & \text{complex surface} \\
L: & \text{special Lagrangian}
\end{array}
\quad \leadsto \quad
\begin{array}{ll}
X_7: & G_2~\text{holonomy} \\
M_4: & \text{coassociative}
\end{array}
\quad \leadsto \quad
\begin{array}{ll}
X_8: & Spin(7)~\text{holonomy}
\end{array}
\label{setup}
\ee
Here, $X$, $S$, and $L$ are in a similar mutual relation as ingredients of open Gromov-Witten theory; namely, $S,L \subset X$ and $L$ has boundary in $S$. As far as we know, moduli spaces of such special Lagrangians with boundary on $S$ have not been studied in the previous literature.\footnote{Surprisingly, even the linearized version of the problem \cite{MR1955286} does not seem to have been pursued.}

In our models, we allow $X$ and $S$ to be compact or non-compact, whereas $L$ is always assumed to be compact with non-trivial boundary
\be
\Sigma := \partial L \; \subset \; S
\ee
In general, we allow $L$ to have several connected components $L_i$, $i = 1, \ldots, \ell$, possibly taken with multiplicity $N_i \in \Z_+$.

\noindent
Examples of $(X,S,L)$ include the following natural choices:

\begin{itemize}
	
\item One can take $X= \C^* \times \C^* \times \C^* = T^* (T^3)$ and $L = T^3$ split into
several ``disks'' $L_i \cong D^3$ (each of which has topology of a 3-ball) by a holomorphic surface $S \subset X$.
These are the so-called brane brick models (BBMs) \cite{Franco:2015tya,Franco:2016nwv,Franco:2016qxh}.

\item More generally, one can take $X = T^* L$ to be the total space of the cotangent bundle over
a 3-dimensional base manifold $L$, with a simple choice of $L$ such as $\R^3$, $T^3$, $S^3$, $S^2 \times S^1$, {\it etc.} Each of the resulting spaces $T^* L$ admits a complete Calabi-Yau metric.

\item The previous class of examples contains the simplest Calabi-Yau 3-fold $X = \C^3$. Even with this choice of $X$, one can generate many interesting models by taking different hypersurfaces $S$ defined by the zero locus of a polynomial $P(z_1,z_2,z_3)$:
\be
S : \qquad P(z_1,z_2,z_3) = 0\,, \qquad (z_1,z_2,z_3) \in \C^3
\label{SPinC3}
\ee

\item Note, same $\Sigma$ can bound several Lagrangians which, moreover, can be special Lagrangian. In 2d $\CN=(0,2)$ theory, different choices of $L$ bounded by the same $\Sigma$ correspond to different (classical) vacua of the same 2d theory:
\be
\CM \left( L; \;
\stackrel{\text{fixed}}{[\Sigma], S} \right) \; \cong \; \CM_{\text{vacua}}
\label{UVvacua}
\ee
For example, let $X$ be the ``deformed conifold'' defined by a quadratic equation
\be
X : \qquad z_1^2 + z_2^2 + z_3^2 + z_4^2 = \epsilon\,, \qquad (z_1,z_2,z_3,z_4) \in \C^4
\ee
A hypersurface defined by imposing an additional equation $z_4=0$ has topology $S \cong T^* S^2$ with a single non-trivial 2-cycle $\Sigma \cong S^2$ that bounds two special Lagrangian disks $L_1 \cong L_2 \cong D^3$ in $X$.

\item Let $I_i \subset \R^3$ and $C_j \subset \R^3$ be a collection of intervals $I_i$ and 2-planes $C_j$, such that all intervals are parallel to each other, with end-points on 2-planes, and such that $C_j$'s are also mutually parallel and orthogonal to $I_i$'s. Every such arrangement defines $X=\C^2 \times T^2$, $S = \sqcup_j (C_j \times T^2)$, and $L = \sqcup_i (I_i \times T^2)$ which corresponds to 2d compactification of the models studied in~\cite{Witten:1997sc}.

\item Let $X$ be the famous quintic Calabi-Yau 3-fold:
\be
X : \qquad f(z_i) = 0\,, \qquad (z_1, \ldots , z_5) \in \cp^4
\ee
where $f(z_i)$ is a homogeneous polynomial of degree 5. Imposing an extra linear relation\footnote{which, without loss of generality, we can assume to be $z_5 = 0$} cuts out a degree-5 surface of general type, which is simply-connected and has $b_2^+ (S) = 9$, $b_2^- (S) = 44$. Since $b_2 (X) = 1$, there are many primitive\footnote{A class in $H^2 (S;\Z) \cong H_2 (S;\Z)$ is called primitive if its wedge product (or, equivalently, a contraction) with the K\"ahler form vanishes. This condition is necessary for $\Sigma$ to be a Lagrangian 2-manifold in $S$.} homology classes in the 53-dimensional lattice $H_2 (S;\Z)$ that are trivial in $X$, thus providing many potential candidates for $\Sigma$ that bound Lagrangian ``disks'' $L \subset X$.

\end{itemize}

While the input data for each of our models is the triple $(X,S,L)$, the output can be summarized either as a pair $(X_7, M_4)$ of a $G_2$ holonomy 7-manifold $X_7$ with a coassociative submanifold $M_4$, or as an 8-manifold $X_8$ of $Spin(7)$ holonomy. They are produced from the input data via the map $\leadsto$ that we call ``M-theory lift,'' because $(X_7,M_4)$ is the result of lifting to eleven dimensions the following type IIA brane configuration,
\be
\begin{matrix}
	{\mbox{\rm space-time:}} & \qquad & \R^4 & \times & X \\
	& \qquad & \cup &  & \cup \\
	{\mbox{\rm NS5-branes:}} & \qquad & \R^2 & \times & S \\
	{\mbox{\rm D4-branes:}} & \qquad & \R^2 & \times & L
\end{matrix}
\label{D4NS5}
\ee
whereas $X_8$ is similarly produced \cite{Gukov:2001hf,Gukov:2002zg} from D6-branes supported on $\R^3 \times M_4$ in type IIA space-time $\R^3 \times X_7$. This M-theory lift of D6-branes supported on $\R^3 \times M_4$ is an auxiliary problem from the viewpoint of our main setup \eqref{D4NS5} and is modeled after \cite{Atiyah:2000zz,Atiyah:2001qf}, where analogous M-theory lift of D6-branes on special Lagrangian 3-manifolds was considered.

When the Calabi-Yau 3-fold $X$ is compact and has irreducible $SU(3)$ holonomy group,
a compactification of type IIA string theory on $X$ produces a 4d $\CN=2$ effective
supergravity theory coupled to matter.
In this theory, a configuration of NS5 and D4-branes as in \eqref{D4NS5} engineers
a $\frac{1}{4}$-BPS surface operator, which preserves $\CN=(0,2)$ supersymmetry on its two-dimensional world-volume $\R^2$.
Since the M-theory lift of this configuration has to preserve the same symmetry and supersymmetry,
we quickly learn that it is given by M5-branes wrapped on a coassociative cycle
in a $G_2$ holonomy manifold,
\be
\begin{matrix}
	{\mbox{\rm space-time:}} & \qquad & \R^4 & \times & X_7 \\
	& \qquad & \cup &  & \cup \\
	{\mbox{\rm M5-branes:}} & \qquad & \R^2 & \times & M_4
\end{matrix}
\label{M5branes}
\ee
When a certain anomaly discussed in section~\ref{sec:anomaly} vanishes,
the topology of the $G_2$ manifold is simply $X_7 \cong S^1 \times X$, and\footnote{
Note, when $L = S^3 \setminus K$ this operation is nothing but the standard knot surgery in $S$ \cite{MR1650308}. See {\it e.g.} \cite{Gadde:2013sca} for physics-friendly introduction to various elements
of 4-manifold topology that will be useful to us here.}
\be
M_4 \; = \; \left( S \setminus \Sigma \right) \; \cup_{\Sigma \times S^1} \; \left( L \times S^1 \right)
\label{M4viaSL}
\ee
When the anomaly of section~\ref{sec:anomaly} is non-trivial, the story involves additional ingredients and the geometry of $X_7$ and $M_4$ becomes more interesting. In particular, $X_7$ is topologically no longer a product $S^1 \times X$, but rather a non-trivial circle fibration:
\be
\begin{matrix}
S^1 & \longrightarrow & X_7 \\
& & \downarrow \\
& & X
\end{matrix}
\label{XXfibration}
\ee
Similarly, $X_8$ is a circle fibration over $X_7$ or, equivalently, a torus fibration over $X$:
\be
\begin{matrix}
	T^2 & \longrightarrow & X_8 \\
	& & \downarrow \\
	& & X
\end{matrix}
\label{XTfibration}
\ee
Upon the lift to $(M_4,X_7)$ or to $X_8$, the moduli problem for Lagrangian ``disks'' $L$ with boundary on $S \subset X$ translates into a more tractable and better understood problem of moduli of $Spin(7)$ metrics or moduli of coassociative submanifolds in $G_2$ holonomy spaces.
We hope that families of new $Spin(7)$ metrics on \eqref{XTfibration} labeled by $(X,S,L)$ can be constructed using methods similar to those in \cite{Foscolo:2017vzf,Foscolo:2018mfs}, where many new $G_2$ analogues of Taub-NUT spaces were found.

The paper is organized as follows. We start in section \ref{sec:example} with a simple example of the D4-NS5 brane model \eqref{D4NS5}.
In section \ref{sec:anomaly} we describe a new chiral anomaly on 4d boundary of D4-branes.
In section \ref{sec:ODE} we construct new coassociative submanifolds in $\R^3 \times \text{Taub-NUT}$ which are ALF generalizations of the ALE coassociatives in $\R^7$ constructed by Harvey and Lawson~\cite{MR666108}.
In section \ref{sec:physics} we analyze in detail the physics of one of the D4-NS5 brane models \eqref{D4NS5} and make a proposal for the 2d $\CN=(0,2)$ effective theory. In particular, via lift to M-theory on a $G_2$ manifold, it gives us a concrete description of the moduli space \eqref{UVvacua} in that brane model:
\be
\CM \left( L; [\Sigma], S \right) \; \cong \; \CM_{\text{vacua}} \; = \; \{ (z,w) \in \C^2 \; \vert \; zw = 0 \}
\label{MvacTNex}
\ee


\section{A simple instructive example}
\label{sec:example}

If one wants to construct the simplest model labeled by $(X,S,L)$,
for the choice of $X$ nothing can be simpler than $\mathbb{C}^3$.
For the choice of complex surface $S$,
it is natural to take the zero locus \eqref{SPinC3} of a polynomial $P(z_1,z_2,z_3)$.
Since we want $S$ to contain at least one non-trivial 2-cycle $\Sigma$,
the degree of $P(z_1,z_2,z_3)$ has to be at least 2.
So, in our simple toy model we make a symmetric choice
\be
S : \qquad P(z_1,z_2,z_3) = z_1^2 + z_2^2 + z_3^2 - \epsilon^2 = 0\,, \qquad (z_1,z_2,z_3) \in \C^3
\label{NS5-locus}
\ee
which has the added benefit of an extra symmetry $SU(2) \cong SO(3)$.
Without loss of generality, we can take the constant $\epsilon$ to be real and positive. 
Topologically, $S$ is a line bundle $\CL = \CO (-2)$ over $\cp^1$.
In particular, $S$ admits a non-contractible 2-cycle $\Sigma \cong S^2$ which, of course,
is contractible in the ambient space $X = \C^3$.
The 2-cycle $\Sigma$ can be described
explicitly by the same equation \eqref{NS5-locus} with $(z_1,z_2,z_3)$ restricted to the real slice:
\be
\Sigma \cong S^2 \; : \; \qquad 
\vec{x}^2 = \epsilon^2 \,, \quad
\vec{y} = 0 \,, \quad 
\vec{x}, \vec{y} \in \mathbb{R}^3 \quad 
(z_k = x_k + i y_k)
\label{Sigmaexample}
\ee
In the ambient space $X = \C^3$ it can be ``filled in'' by a disk --- or, rather, by a 3-dimensional ball ---
with boundary $\Sigma$:
\be
L \;\; : \qquad 
\vec{x}^2 \le \epsilon^2 \,, 
\quad
\vec{y} = 0
\label{D4-disk}
\ee
It is easy to check that $L$ is, in fact, special Lagrangian with respect to the flat Calabi-Yau structure on $X = \C^3$:
\be
\Omega = dz_1 \wedge dz_2 \wedge dz_3
\qquad , \qquad
J \; = \; \frac{i}{2} dz_1 \wedge d \bar z_1
+ \frac{i}{2} dz_2 \wedge d \bar z_2
+ \frac{i}{2} dz_3 \wedge d \bar z_3
\label{CYflat}
\ee

Once we introduced the key players, $X$, $S$, and $L$,
they appear to define a perfectly sensible configuration of NS5 and D4-branes \eqref{D4NS5}
with world-volumes $\R^2 \times S$ and $\R^2 \times L$, respectively.
This choice of $(X,S,L)$, however, suffers from an important anomaly.
As we explain in the next section, this anomaly can be easily cured.
However, as it stands, the triple $(X,S,L)$ introduced here does not define a consistent
model and, in particular, does not admit a lift \eqref{setup} to a coassociative submanifold $M_4 \subset X_7$
or to a $Spin(7)$ manifold~$X_8$.


\section{Euler number anomaly}
\label{sec:anomaly}

The simple choice of $(X,S,L)$ introduced in the previous section is a good example to
illustrate an important anomaly in our class of models, which in general takes values in
\be\label{homSigma}
H^2 (\Sigma, \Z) \; \cong \; H_0 (\Sigma, \Z)
\ee
Specifically, for each connected component of $\Sigma$, there is a potential $\Z$-valued
anomaly given by the self-intersection of that component in the complex surface $S$:
\be
n_i \; = \; - \Sigma_i \cdot \Sigma_i \,, \qquad i =1, \ldots, \dim H_0 (\Sigma,\Z)
\label{nSS}
\ee
This anomaly can be seen both physically and geometrically. In this section we begin by illustrating the anomaly for the simple example in section 
\ref{sec:example}, focusing on the M-theory geometric perspective. In section \ref{sec:anomalybrane} we then present a general discussion of the anomaly in type IIA string theory, how this lifts to the 
M-theory picture, and how the anomaly can be cured (in a more or less canonical way). Finally, section~\ref{sec:anomalyQFT} presents the QFT manifestation of the anomaly in our class of models.

\subsection{Example}\label{sec:anomalyexample}

In our simple example \eqref{NS5-locus}--\eqref{D4-disk}, the lift to M-theory involves a 7-manifold of $G_2$ holonomy, $X_7$, which topologically is simply a product of $X$ and the M-theory circle $S^1$. (This is a general feature of any type IIA background that does not involve D6-branes or RR 2-form fluxes.) Likewise, a lift of the Lagrangian disk \eqref{D4-disk} is also a product $L \times S^1$, which is automatically coassociative in $X_7 = X \times S^1$ with respect to the natural $G_2$ structure~\eqref{coass-cy}. The 4-manifold $L \times S^1$, however, has boundary $\Sigma \times S^1$ and needs something to end on. Indeed, both NS5 and D4-branes turn into M5-branes upon M-theory lift. And, just like in type IIA theory the NS5-brane \eqref{NS5-locus} serves as a boundary condition for the D4-brane \eqref{D4-disk}, in M-theory they join into one single 4-manifold since the M-theory lift of the NS5-brane is simply $S \subset X_7$.

Now we are starting to see the anomaly or, at least, its geometric manifestation. The boundary of $L \times S^1$ is $\Sigma \times S^1$. The boundary of $S \setminus \Sigma$ locally looks very similar; the normal sphere bundle to $\Sigma \subset S$ is just a circle $S^1$ bundle. However, when the normal bundle is non-trivial, the boundary of $S \setminus \Sigma$ is a non-trivial $S^1$ bundle over $\Sigma$ of degree $n$ given by the self-intersection \eqref{nSS}. Therefore, when $n \ne 0$, we cannot simply glue $S \setminus \Sigma$ and $L \times S^1 \subset X_7$ into a single 4-manifold \eqref{M4viaSL} along their boundaries since these boundaries are different. This is precisely the case in our simple example, where $n=-\Sigma \cdot \Sigma = 2$.

\subsection{Type IIA interpretation and M-theory lift}\label{sec:anomalybrane}

The geometric anomaly just illustrated may be seen from the type IIA brane perspective as follows. 
On the worldvolume of the NS5-brane propagates a periodic scalar field $\varphi$, where we normalize $\varphi$ to have period $2\pi$. This worldvolume theory arises from dimensional reduction of the M5-brane 
theory \cite{Bandos:2000az}, where the NS5-brane is transverse to the M-theory circle direction, $S^1_M$. The scalar $\varphi$ then 
corresponds to motion of the M5-brane in that direction. On the other hand, a D4-brane ending on an NS5-brane acts as 
a magnetic source for this scalar \cite{Witten:1997sc}. In our set-up, recall from \eqref{D4NS5} that the NS5-brane 
worldvolume is $\R^2\times S\subset \R^4\times X$, where the D4-brane wrapped on $\R^2\times L$
shares the $\R^2$ directions with the NS5-brane. Suppressing the common $\R^2$ directions, the end of the D4-brane 
is then $\partial L=\Sigma\subset S$. Since this has codimension 2 in $S$, it is 
linked by a circle $S^1$. 
As one goes around such a transverse $S^1\subset S$, the periodic scalar $\varphi$ winds once around 
$2\pi$, so that $\R^2\times \Sigma$ is effectively the locus of a  ``vortex'' for the $\varphi$ field. 

We may describe this in more detail as follows. The scalar $\varphi$ enters the worldvolume theory of the NS5-brane via its gauge-invariant 
curvature \cite{Bandos:2000az}
\be\label{F1definition}
\CF_1\equiv d\varphi- \iota^*C_1
\ee
where $C_1$ denotes the RR 1-form potential, and $\iota:\R^2\times S\hookrightarrow \R^4\times X$ denotes the embedding 
of the NS5-brane into space-time, with $\iota^*$ a pull-back to the worldvolume. Of course, 
we might {\it a priori} choose to turn off this RR potential, setting $C_1=0$, since neither the NS5-brane nor D4-brane source it. 
To say that $\Sigma\subset S$ is a magnetic source for $\varphi$ then means that 
\be\label{vortexsource}
d\CF_1 = 2\pi \delta_2
\ee
Here $\delta_2$ is a delta-function representative of the Poincar\'e dual of $\Sigma\subset S$, {\it i.e.}
$\delta_2$ is a closed 2-form on $S$  which restricts to a Dirac delta-function times the volume form 
in the normal 
$\R^2$ directions of $\Sigma$. 
Integrating \eqref{vortexsource} along such a normal $\R^2$, with 
positive constant radius circle $S^1\subset \R^2$, one obtains
\be\label{windingvarphi}
2\pi = \int_{\R^2}d\CF_1 = \int_{S^1} d\varphi 
\ee
This says that $\varphi$ winds once around $2\pi$ as it moves around an $S^1$ linking $\Sigma$. 
More generally for $k$ D4-branes the winding is $2\pi k$.

On the other hand, from \eqref{F1definition} we have in general  that
\be\label{BianchiF1}
-\iota^*G_2 = d\CF_1 = 2\pi \delta_2
\ee
where $G_2=dC_1$ is the RR 2-form flux. Equating cohomology classes of the left and right hand sides of \eqref{BianchiF1} then 
says
\be\label{fixG2}
[\iota^* G_2] = -2\pi [\delta_2] \in H^2(S,\R)
\ee
In other words, when the end of the D4-brane $\Sigma\subset S$ has a Poincar\'e dual $\delta_2$ that defines 
a non-trivial cohomology class in $H^2(S,\R)$, one must \emph{necessarily} turn on a RR $G_2$ field 
obeying \eqref{fixG2}: to not do so leads to our anomaly. 

Let us examine this further. We denote the normal bundle of $\Sigma$ inside $S$ as $N\Sigma$, which is a complex line bundle. 
Such line bundles are classified by their first Chern class, which here lies in 
$H^2(\Sigma,\Z)\cong \Z^{b_0(\Sigma)}$, where as in \eqref{homSigma} 
$b_0(\Sigma)=\dim H_0(\Sigma,\Z)$ is the number of connected components of $\Sigma$. 
For the $i$th  component $\Sigma_i$ we may then 
identify $N\Sigma_i =\CO(-n_i)$ as the line bundle over $\Sigma_i$ of Euler number 
$-n_i\in\Z$. More 
abstractly, for each $i=1,\ldots,b_0(\Sigma)$  the Poincar\'e dual
$\delta_2$ represents the generator 
of the compactly supported cohomology $[\delta_2]_{\mathrm{cpt}}=1\in H^2_{\mathrm{cpt}}(N \Sigma_i,\Z)\cong H^0(\Sigma_i)\cong \Z$. 
This generator is also known as the \emph{Thom class}. There is a natural mapping from $\Z\cong H^2_{\mathrm{cpt}}(N\Sigma_i,\Z) 
\rightarrow H^2(N\Sigma_i,\Z)\cong H^2(\Sigma_i,\Z)\cong \Z$ where the  map simply forgets the compact support condition. 
This maps the Thom class $1\in\Z$ to the Euler number $-n_i\in\Z$. Putting all this together, we may integrate
\eqref{BianchiF1} over $\Sigma_i$ to obtain
\be\label{G2ni}
\int_{\Sigma_i} \frac{G_2}{2\pi} = n_i
\ee

Recall that geometrically $G_2$ is the curvature of the M-theory circle bundle over space-time. Thus
\eqref{G2ni} says that we must fiber the M-theory circle bundle in such a way that the 
Euler number of this fibration over $\Sigma_i$, which is the left hand side of \eqref{G2ni}, is the same as the 
self-intersection number $n_i=-\Sigma_i\cdot \Sigma_i$, which is also minus the Euler number 
of the normal bundle $N\Sigma_i$ of $\Sigma_i$ in $S$. There are two ways to generate such a flux in our set-up,
where recall that space-time is $\R^4\times X$, with $X$ a Calabi-Yau 3-fold: we may turn 
on RR 2-form flux on $X$, and/or introduce D6-branes into the space-time, which magnetically 
source such a flux. In order to preserve supersymmetry, the RR 2-form flux on $X$ should be 
Hodge type $(1,1)$, and D6-branes should wrap a special Lagrangian $L_{D6}\subset X$ 
and are space-filling in the $\R^4$ directions. The lift of $X$ to M-theory is then not simply 
a product $X\times S^1_M$, but rather the total space $X_7$ a non-trivial circle fibration over $X$, \eqref{XXfibration}, which degenerates 
at D6-brane loci. Supersymmetry implies that $X_7$ is a $G_2$ holonomy manifold. 
Notice that in the simple case that $X=\C^3$, as in the example in section \ref{sec:example}, 
since $H^2(X,\Z)=0$ we must necessarily introduce D6-branes to source the required flux on the left hand side 
of \eqref{G2ni}. We shall return to discuss this example at the end of the next subsection.

The condition \eqref{G2ni} has a very elegant geometric interpretation, once we lift the NS5-brane and D4-brane 
configuration to M-theory. Both objects descend from an M5-brane, where the NS5-brane is transverse to
the M-theory circle while the D4-brane wraps the circle. Since the D4-brane ends on the NS5-brane, the configuration
should lift to a \emph{single} M5-brane wrapped on a 4-manifold $M_4$ in M-theory. The classical picture of the D4-brane  ending on 
a definite submanifold $\Sigma\subset S$ inside the NS5-brane is not quite accurate, due to the resulting distortion 
near to this locus. Let us examine the lift of both branes away from the intersection locus $\Sigma$. 
The NS5-brane wraps $S\setminus \Sigma$, which near to a connected component $\Sigma_i$ looks like $N\Sigma_i\setminus \Sigma_i\cong 
I\times M_3$, where $I=(0,\epsilon]$ is an interval and $M_3$ is the total space of a degree $n_i$ circle 
bundle over $\Sigma_i$
\be
M_3 \quad = \quad
\begin{matrix}
	S^1 & \\
	\downarrow & \text{deg.}~n_i \\
	\Sigma_i & 
\end{matrix}
\label{M3viaSSigma}
\ee
On the other hand, the D4-brane wraps $L\setminus \Sigma$, which near 
to the $i$th end looks like $I\times \Sigma_i$. On lifting to M-theory, the M-theory circle 
fibres over this latter geometry to give the M5-brane worldvolume, and \eqref{G2ni} says 
that the total space of this $S^1_M$ bundle over $I\times \Sigma_i$ is also precisely $I\times M_3$. 
The condition \eqref{G2ni} thus ensures that the boundaries of these two  neck regions, around where the NS5-brane and D4-brane meet, 
lift to the same 3-manifold $M_3$ in M-theory, and as in  \cite{Witten:1997sc} these are glued together 
to produce a single smooth M5-brane by connecting the necks via a small ``tube'' $[-\epsilon,\epsilon]\times M_3$. 
The M5-brane is then wrapped on the smooth 4-manifold
\be
M_4 = (S\setminus \Sigma)\cup_{M_3}(S^1_M\rightarrow L\setminus \Sigma)
\ee
On the other hand, without the addition 
of the RR 2-form flux the boundaries are different 3-manifolds, so we cannot simply glue them together. 
We saw this for our explicit example in section \ref{sec:anomalyexample}.

We conclude this subsection with a few more general remarks. Suppose one 
has a submanifold $S$ in space-time over which there is a non-trivial 
RR 2-form flux, {\it i.e.} the pull-back $[\iota^*G_2]\neq 0 \in H^2(S,\R)$. 
Then one \emph{cannot} wrap an NS5-brane over $S$. Geometrically, 
this is because the NS5-brane is an M5-brane transverse to the M-theory circle, 
and thus the M-theory circle bundle over $S$ must have a global section. But 
this is true if and only if the $S^1_M$ bundle is trivial over $S$, and hence $[\iota^*G_2]= 0 \in H^2(S,\R)$. 
From the point of view of the NS5-brane worldvolume theory, one sees the same 
fact from the equation $d\CF_1=-\iota^*G_2$. Since the worldvolume periodic scalar $\varphi$ 
is also a section of the same M-theory circle bundle over $S$, this scalar field exists 
as a global function on $S$ if and only if that bundle is trivial. In this case $\CF_1=d\varphi-\iota^*C_1$
is a global 1-form on $S$, and hence $\iota^*G_2$ is exact. If instead $[\iota^*G_2]\neq 0 \in H^2(S,\R)$, 
the best one can do is to remove a submanifold $\Sigma\subset S$ that is Poincar\'e dual 
to $-[\iota^*G_2/2\pi]$. By construction, the M-theory circle bundle is then trivial over $S\setminus\Sigma$, 
and an NS5-brane can wrap this locus. Of course, physically we may then require a D4-brane 
to end on the locus $\Sigma$, thus satisfying \eqref{BianchiF1}, and this system then lifts to a single smooth M5-brane, 
without additional boundaries. In this case the worldvolume scalar $\varphi$ is not defined at the locus $\Sigma\subset S$ 
due to the winding by $2\pi$ in \eqref{windingvarphi}, just as the angular coordinate $\theta$ is 
not defined at the origin of $\R^2$ in plane polar coordinates.

Similar remarks apply to a D2-brane with a fundamental string ending on it, which lifts to an M2-brane. 
In particular, the D2-brane worldvolume theory contains a periodic scalar corresponding to motion 
of the M2-brane in the M-theory circle direction. In three dimensions a periodic scalar is dual 
to an Abelian gauge field, which in this case is nothing but the usual $U(1)$ gauge field on a D-brane. 
A fundamental string ending on a D-brane is of course an electric source for this gauge field, 
or equivalently for the D2-brane a magnetic source for the dual periodic scalar. 

\subsection{QFT interpretation}
\label{sec:anomalyQFT}

Now let us discuss the QFT origin and interpretation of this ``Euler number anomaly.'' Just like in its geometric manifestation, the crucial ingredient is the neck region along which $S \setminus \Sigma$ and $L \times S^1$ are supposed to be joined. Recall from the previous subsection that near the $i$th connected component of this neck region, 
$S \setminus \Sigma$ looks like $I \times M_3$, where $I = (0,\epsilon]$ and  $M_3$ is the total space of a degree $n_i$ circle bundle over $\Sigma_i$ 
in \eqref{M3viaSSigma}.
Let us look more closely at the physics of a five-brane near this neck region, {\it i.e.} a five-brane on $\R^2 \times I \times M_3$. More generally, we can consider $N$ five-branes supported on $\R^2 \times I \times M_3$.

\begin{figure}[ht]
	\centering
	\includegraphics[width=2.7in]{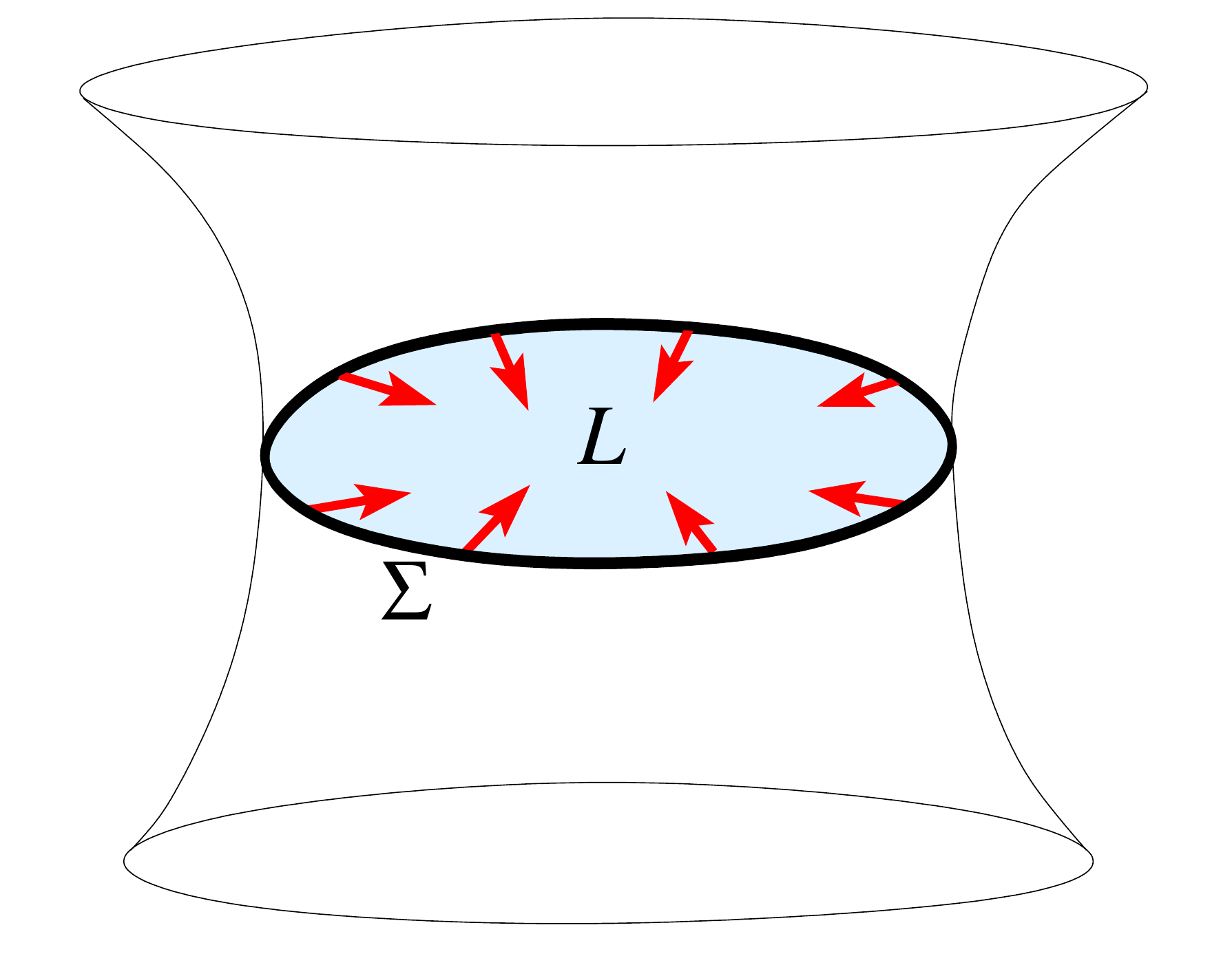}
	\caption{An illustration of the anomaly inflow from the boundary into the bulk of the D4-brane world-volume theory.}
	\label{fig:anomaly1}
\end{figure}

Since $I \times M_3$ is part of the curved 4-manifold along which the five-brane is topologically twisted, a priori the interval $I$ is not quite on the same footing as the $\R^2$ part of the five-brane world-volume. However, since $I$ admits a flat metric, and since in flat space the full physical and topological theories are the same, we can treat $\R^2 \times I$ as a space-time of the 3d physical theory obtained by compactifying a five-brane on $M_3$.

This physical 3d theory has $\CN=2$ supersymmetry and is usually denoted $T[M_3]$. It is defined for any 3-manifold $M_3$, but in our applications here we only need to know this theory for very special 3-manifolds of the form \eqref{M3viaSSigma}. For example, when $\Sigma = S^2$, as in \eqref{Sigmaexample}, we have $M_3 = L(n,1)$ and $T[M_3,G]$ is the following ``Lens space theory'' \cite{Acharya:2001dz,Gadde:2013sca}:
\be
T[L(n,1),G]
\quad = \quad
\boxed{\begin{array}{c}
		~\text{3d } \CN=2 \text{ super-Chern-Simons} \\
		~\text{with } G_n \text{ and an adjoint chiral}
\end{array}}
\label{TLens}
\ee
The important aspect of this theory is that it involves  a Chern-Simons coupling for the gauge group $G$ at level $n$.

In our simple example \eqref{D4-disk}, the interval $I = (0,\epsilon]$ is parametrized by the radial coordinate $r = |\vec x|$, and the 3d theory we described is a result of the reduction of the five-brane world-volume theory on the M-theory circle and $\Sigma = S^2$ (or, more precisely, on an $S^1$ bundle over $\Sigma$). At the end-points of the interval $I$ we need to impose 2d boundary conditions, which must preserve $\CN=(0,2)$ supersymmetry because this is the amount of supersymmetry preserved by the brane configuration.

\begin{figure}[ht]
	\centering
	\includegraphics[width=2.7in]{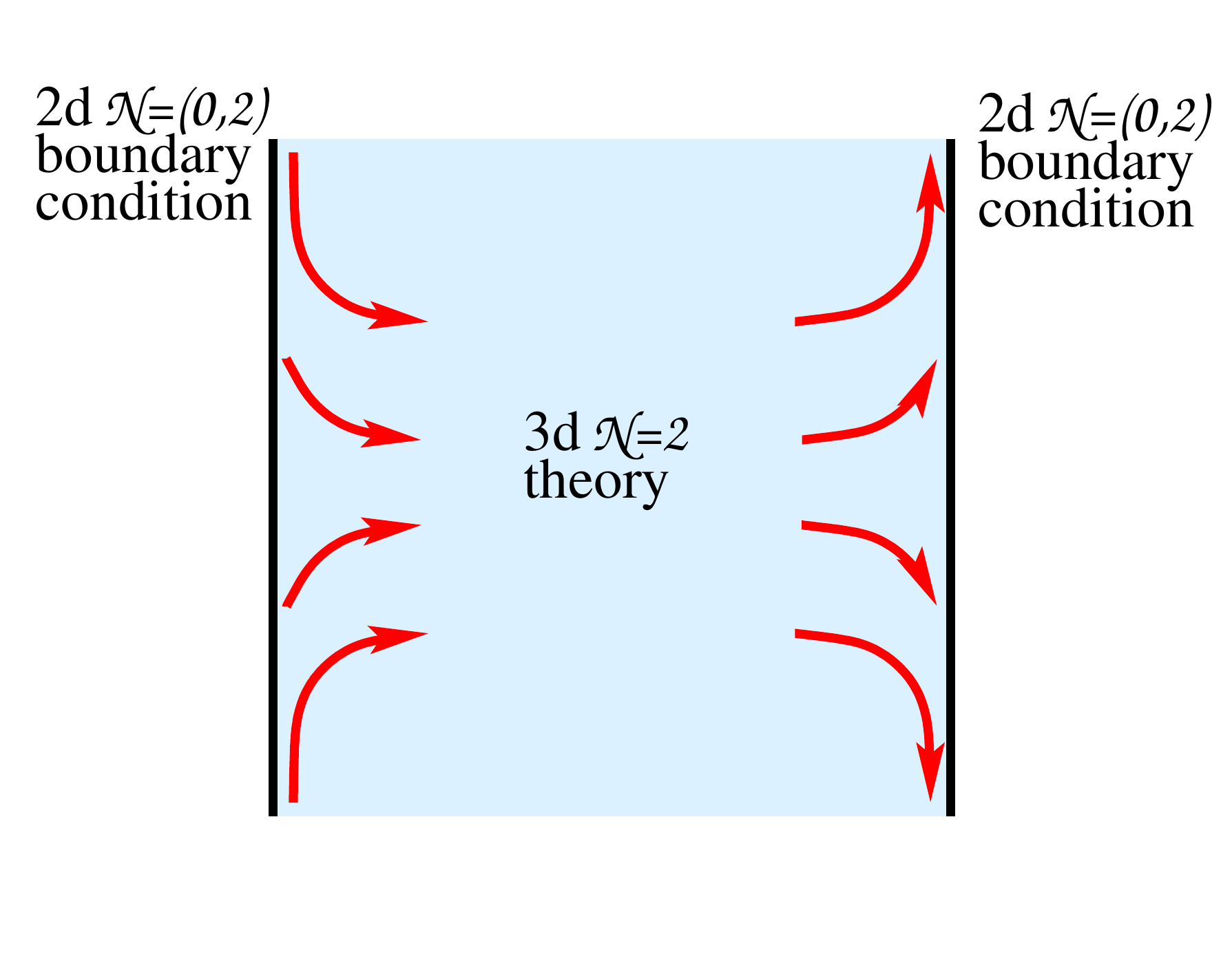}
	\caption{An illustration of the anomaly inflow in 3d $\CN=2$ theory with 2d $\CN=(0,2)$ boundary conditions. For example, if one boundary carries charged chiral fermions whose anomaly is compensated by Chern-Simons couplings of the 3d bulk theory, a similar mechanism must be at work at the other boundary.}
	\label{fig:anomaly2}
\end{figure}

The 2d $\CN=(0,2)$ boundary conditions in question have a geometric origin: at one end of the interval the boundary condition is determined by the 4-manifold $S \setminus \Sigma$. This is where D4-brane runs into NS5-brane. At the other end of the interval, the boundary condition is determined by what happens at the limit $r=0$. The resulting system looks like a 3d $\CN=2$ theory \eqref{TLens} sandwiched by 2d $\CN=(0,2)$ boundary conditions, precisely of the type studied in \cite{Gadde:2013sca} and illustrated in Figure~\ref{fig:anomaly2}.

In particular, when $n \ne 0$, there is a non-trivial anomaly inflow that also played an important role in \cite{Gadde:2013sca}. And, what our above analysis shows is that the boundary condition where the D4-brane runs into the NS5-brane infuses $n$ units of anomaly compensated by the Chern-Simons coupling of the 3d theory and also carried to the other boundary. Therefore, the other boundary condition (at $r=0$ in our simple example) should also carry $n$ units of anomaly, {\it i.e.} it should carry chiral degrees of freedom charged under $G$ (= gauge group of D4-brane theory).

The original system of NS5 and D4-branes has no such degrees of freedom at $r=0$. Therefore, it is anomalous. However, our analysis in the previous subsection also suggests what one should do in order to cancel this anomaly. The simplest possibility is to add $n$ D6-branes that, geometrically, would replace a product with the M-theory circle by a non-trivial bundle and, physically, would produce $n$ charged chiral fermions at the 2-dimensional intersection of the D4 and D6-branes (the lowest modes of D4-D6 open strings). As mentioned earlier, in general another possibility is to turn on RR 2-form flux. In this paper, we mostly consider the first option.

The extra D6-branes that we need to add in order to cancel the Euler number anomaly (when $n \ne 0$) should have the following properties. First, their 7-dimensional world-volume should intersect the world-volume $\R^2 \times L$ of the D4-branes along $\R^2$ (times a point in $L$, which in our simple example is best to be chosen at $r=0$ in order to preserve $SU(2)$ symmetry of the background). Second, the D6-branes must preserve the unbroken supersymmetry of the original brane system~\eqref{D4NS5}. These two conditions imply that D6-branes must be supported on $\R^4 \times L_{D6}$, where $L_{D6} \subset X$ is a special Lagrangian submanifold calibrated by the 3-form $\text{Im} (\Omega)$, provided $L$ is calibrated by $\text{Re} (\Omega)$:
\be
\begin{array}{ccccl}
	{\mbox{\rm space-time:}} & \qquad & \R^4 & \times & X \\
	\hline
	{\mbox{\rm NS5-branes:}} & \qquad & \R^2 & \times & S \\
	{\mbox{\rm D4-branes:}} & \qquad & \R^2 & \times & L \\	
	{\mbox{\rm D6-branes:}} & \qquad & \R^4 & \times & L_{D6}
\end{array}
\label{D4NS5D6}
\ee
See Appendix~\ref{sec:D6} for details.

In our model example from section \ref{sec:example}, a special Lagrangian $L_{D6} \subset \C^3$ that meets \eqref{D4-disk} at $\vec x = 0$ and is calibrated by $\text{Im} (\Omega)$ is a three-dimensional plane:
\be
L_{D6} \;\; : \qquad 
\vec{y} \in \R^3 \,, 
\quad
\vec{x} = 0
\label{LD6R3}
\ee
In its presence, the corresponding M-theory lift of \eqref{D4NS5D6} is of the form \eqref{XXfibration}:
\be
X_7 \; = \; \R^3 \times \text{TN}_n
\label{XTNn}
\ee
with a coassociative submanifold, {\it cf.} \eqref{M4viaSL},
\be
M_4 \; = \; \left( S \setminus \Sigma \right) \; \cup_{M_3} \; \text{Cone} (M_3)
\label{generalM4M3}
\ee
where $M_3$ is the 3-manifold introduced in \eqref{M3viaSSigma}. Since in our example the complex surface $S$ is itself asymptotically a cone on $M_3$, and $n=2$, it follows that $M_4 = \text{Cone} (M_3) \cong \R^4 / \Z_2$.

Note, instead of D4-branes supported on Lagrangian disks \eqref{D4-disk}, illustrated in Figure~\ref{fig:anomaly1}, one can consider non-compact D4-branes supported on the ``disk complement,''
\be
L \;\; : \qquad 
\vec{x}^2 \ge \epsilon^2 \,, 
\quad
\vec{y} = 0
\label{diskcompl}
\ee
Although this option may seem less natural for the purpose of building a dynamical 2d gauge theory on D4-brane world-volume, it does have very interesting physics, as will be discussed in section~\ref{sec:physics}. Brane models with such non-compact D4-branes also exhibit the familiar Euler number anomaly which, much like \eqref{nSS}, can be expressed in terms of the D4-brane boundary components $\Sigma_i \subset S$. However, since the orientation of the boundary of a D4-brane on a ``disk complement'' is reversed compared to that supported on a Lagrangian disk, the Euler number anomaly in these two cases differs by sign,
\be
n_i \; = \; + \, \Sigma_i \cdot \Sigma_i
\label{nSSsign}
\ee
which is equivalent to $n_i \to - n_i$. In our simple class of models with $n$ D6-branes, replacing compact D4-branes on Lagrangian disks by non-compact D4-branes supported on disk complements \eqref{diskcompl} leads to coassociative submanifold in \eqref{XTNn} that are double-ended cones on
\be
S^3/\Z_n \; \sqcup \; - \, S^3/\Z_n
\ee
When the apex is smoothed out, these become topologically \be
M_4 \; \cong \; \R \times S^3/\Z_n
\ee
The next section offers an explicit construction of such coassociative submanifolds.


\section{New coassociative submanifolds}
\label{sec:ODE}

When $X_7 = S^1 \times X$, the associative and coassociative forms on $X$ are
\begin{eqnarray}
\Phi & = & J \wedge d\psi - \text{Im}(\Omega)  \label{coass-cy} \\
\Psi & = & \frac{1}{2} J \wedge J + \text{Re}(\Omega) \wedge d\psi \nonumber
\end{eqnarray}
where $\psi$ is a coordinate on $S^1$.
In particular, it is clear that, given a special Lagrangian $L \subset X$ calibrated by $\text{Re}(\Omega)$ and a complex surface $S \subset X$, both $S^1 \times L$ and $S$ are individually coassociative in $X_7$.

We are interested in a coassociative 4-manifold $M_4 \subset X$ defined by ``smoothing'' of $S^1 \times L$ and $S$ near the ``neck'' region $S \cap ( L \times S^1) = S^1 \times \Sigma$. As we saw in section~\ref{sec:anomaly}, such smoothing is possible only if the self-intersection of $\Sigma$ inside $S$ vanishes. If that is the case, $M_4$ is given by the (deformation of) eq. \eqref{M4viaSL}. When the self-intersection of $\Sigma$ in $S$ is non-zero, we need to replace $X_7 = S^1 \times X$ by a non-trivial circle bundle \eqref{XXfibration}, such that
\be
X_7 / S^1 \; \cong \; X
\ee

Motivated by our discussion in sections \ref{sec:example} and \ref{sec:anomaly}, let us analyze more carefully how this works in the class of examples with $X = \C^3$ and one component \eqref{LD6R3} of the codimension-4 fixed point set supported on $\R^3 \subset \C^3$. For a circle bundle of degree $n$, the total space \eqref{XTNn}
\be
X_7 \; = \; \Lambda^{2,+} (\text{TN}_n) \; \cong \; \R^3 \times \text{TN}_n
\label{XTaubNUT}
\ee
carries a $G_2$ structure determined by the first relation,
\be
\Phi \; = \; dy_1 \wedge dy_2 \wedge dy_3 - \sum_{i=1}^3 \omega_i \wedge dy_i
\label{Phiweneed}
\ee
where $\omega_i$ are self-dual 2-forms on the Taub-NUT space $\text{TN}_n$ of ``charge'' $n$.
The standard way to write a hyper-K\"ahler metric on the Taub-NUT space, which is ideally suited for our application to \eqref{XTNn} with coordinates $(\vec x, \vec y, \psi)$, is
\be
ds^2(\text{TN}_n) = H d\vec{x} \cdot d \vec x + H^{-1}(d\tilde{\psi} + \vec \chi\cdot d\vec{x})^2
\label{metric-TN}
\ee
where $H (\vec x)$ is the harmonic function on $\mathbb{R}^3 \cong \text{TN}_n/S^1$, such that
\be
\nabla \times \vec \chi = \nabla H \,.
\ee
Equivalently, $d\chi = *_3 dH$, where $*_3$ is the Hodge dual with respect to the flat metric of $\mathbb{R}^3$ parametrized by $\vec x$.
In these conventions, the triplet of the self-dual 2-forms on $\text{TN}_n$ that appear in \eqref{Phiweneed} can be explicitly written as
\be
\vec \omega = (d \tilde \psi + \vec \chi \cdot d \vec x)\wedge d\vec{x} - \frac{1}{2} H (d \vec x \times d \vec x) \,,
\ee
where $(d \vec x \times d \vec x)^i = \epsilon^{ijk} dx^j \wedge dx^k$.

In order to preserve the $SO(3)$ symmetry of the triple $(X,S,L)$ in our main example \eqref{NS5-locus}--\eqref{D4-disk}, let us choose the harmonic function $H$ in the form
\be
H = B + \frac{n}{|\vec x|} =  B + \frac{n}{r} \,, 
\label{Hchoice}
\ee
where the constant $B$ is related to the ratio between the radius of the M-theory circle and the eleven dimensional Planck scale. 
Then, the expression \eqref{Phiweneed} is invariant under the $SO(3)$ symmetry, which acts on $\omega_i$ and $dy_i$ in the same way.

Note, all of our ingredients in the original setup \eqref{NS5-locus}--\eqref{D4-disk} can be easily described in the coordinates $\vec x$ and $\vec y$. Namely, the complex surface \eqref{NS5-locus} looks like
\be
S \; : \qquad
\vec{x}^2 - \vec{y}^2 = \epsilon^2 \,,
\qquad 
\vec{x}\cdot \vec{y} = 0 \,.
\label{D4NS5a}
\ee
and $\Sigma \cong S^2$ was already described in \eqref{Sigmaexample} as a sphere in $\vec x$-plane of radius $\epsilon$. The Lagrangian $L$ is a 3-ball (``disk'') within this plane, whereas the Lagrangian $L_{D6} \cong \R^3$ is a copy of $\vec y$-plane at $\vec{x} = 0$, {\it cf.} \eqref{LD6R3}.

In order to construct new coassociative 4-manifolds $M_4$ in \eqref{XTaubNUT} with the $SO(3)$ symmetry, we need to fix the angular dependence as in \eqref{D4NS5a} and allow more general radial dependence that will be encoded in a single function $g$ of the radial variable $r = |\vec{x}|$. In particular, separating radial and angular variables, the equation for complex surface \eqref{D4NS5a} can be written as
\be
\vec{x} = \epsilon (\cosh\rho) \hat{m} \,,
\quad 
\vec{y} = \epsilon (\sinh\rho) \hat{n} \,,
\quad 
|\hat{m}| = |\hat{n}| = 1\,,
\quad 
\hat{m} \cdot \hat{n} = 0 \,.
\label{D4NS5b}
\ee
Note, the locus of small constant $\rho$ defines a circle bundle over $\Sigma = S^2$.

Then, inspired by \eqref{D4NS5a} and \eqref{D4NS5b}, we look for a coassociative 4-manifold $M_4$ of the form
\be
\vec{y} = g(r) \hat{n}(\theta, \phi,\psi) \,,
\qquad
\hat{n} \cdot \hat{n} = 1\,,
\qquad
\hat{m} \cdot \hat{n} = 0 \,.
\label{ansatz-y}
\ee
where $\hat{m} = (\theta,\phi,\psi)$ denote the standard angle coordinates on $S^3$. In these coordinates, the explicit expression for 1-form $\chi$ compatible with our choice of the harmonic function \eqref{Hchoice} is $\chi =  n \cos\theta d\phi$, and the Taub-NUT space is parametrized by the ``polar'' coordinates $(r,\theta,\phi,\psi)$ instead of the original ones $(\vec{x},\tilde{\psi})$.
The angular coordinate $\psi$ here is related to $\tilde{\psi}$ in \eqref{metric-TN} by 
\be
\tilde{\psi} = n \psi  \,. 
\ee

{}From the last two equations in \eqref{ansatz-y} it follows that, for each given $\hat{m}$, the unit vector $\hat{n}$ takes values in a circle. In fact, here we are interested in a solution such that $\hat{n}$ winds exactly once around this circle as $\psi$ runs from $0$ to $2\pi$. Implementing this in our ansatz \eqref{ansatz-y} and requiring that $M_4$ is calibrated with respect to \eqref{Phiweneed}--\eqref{Hchoice}, we obtain a single ordinary differential equation (ODE) for the function $g(r)$:
\be
\boxed{ \phantom{\oint^{\int}_{\int}} \frac{dg}{dr} = \frac{(2n+Br)g}{g^2-nr} \phantom{\oint^{\int}_{\int}}} \,.
\label{ode-final}
\ee
In fact, as we explain momentarily, the specific value of $B$ is irrelevant, as long as $B \ne 0$. It can be changed to any other value by rescaling $r$ and $g$. Therefore, as far as the dependence on $B$ is concerned, there are essentially two cases to consider: $B=0$ and $B \ne 0$.

The ordinary differential equation \eqref{ode-final} is a special instance (with $A=4$) of a more general family of ODEs
\be
\frac{dg}{dr} = \frac{(A+Br)g}{g^2-2r} 
\label{ode-AB}
\ee
which are equivalent, via $f = \frac{1}{2} g^2 - r$, to the Abel equation of the second kind\footnote{We thank Robert Bryant for pointing this out to us.}
\be
f \frac{d f}{dr} = (A-1+Br) f + r (A +Br)
\label{Abelsecond}
\ee
The substitution $z = \int (A-1+Br) dr$ brings it to the canonical form\footnote{Also note that, via a change of variable $F = f^{-1}$, we can bring \eqref{Abelsecond} into the form of the Abel equation of the first kind:
$$
\frac{d F}{dr} = - (A-1+Br) F^2 - r (A +Br) F^3
$$}
\be
f \frac{d f}{dz} = f + \Phi (z)
\ee
where $\Phi(z)$ is defined parametrically, by eliminating $r$ in the relations
\be
\Phi = \frac{Ar + Br^2}{A-1+Br}
\,, \qquad
z = (A-1)r + \frac{B}{2} r^2 + \text{const}
\ee

In analyzing the solutions of \eqref{ode-final} and \eqref{ode-AB} it will be convenient to consider even a larger family of ODEs:
\be
\frac{dg}{dr} = \frac{(A+Br)g}{g^2-2Cr}
\label{ODE-ABC}
\ee
with three parameters, $A$, $B$ and $C$.
When parameters $A$ and $C$ are related as in \eqref{ode-final}, {\it i.e.} $A = 2n = 4C$, this more general ODE has a peculiar property that all parameters can be scaled away completely\footnote{Specifically, under \eqref{grscaling} this equation becomes
$$
\frac{dg}{dr} = \frac{(4 \lambda_g^{-2} \lambda_r C + B \lambda_g^{-2} \lambda_r^2 r)g}{g^2 - 2 \lambda_g^{-2} \lambda_r Cr}
$$} by rescaling $g$ and $r$,
\be
g \; \to \; \lambda_g g
\,, \qquad
r \; \to \; \lambda_r r
\label{grscaling}
\ee
In particular, when $B \ne 0$, without loss of generality we can bring it to a form \eqref{ode-AB} with $A=4$ and $B=1$.

Now, let us explore solutions to the above Abel equation, starting with the special ones that correspond to setting either $A$ or $B$ to zero. When $n=0$, {\it i.e.} $A=C=0$ in \eqref{ODE-ABC}, a simple solution is
\be
\text{NS5-brane}: \qquad g^2 - B r^2 \; = \; \text{const}
\label{NS5sol}
\ee
It describes the original NS5-brane \eqref{D4NS5a} without D6 or D4-branes. Another special case corresponds to setting $B=0$ in \eqref{ode-AB} and leads to a famous asymptotically conical coassociative submanifold in $\R^7$ constructed by Harvey and Lawson~\cite{MR666108}.

Indeed, after a change of variable $r = \rho^2 / 8$ our ODE with $B=0$ has the property that both the numerator and denominator on the right-hand side are homogeneous (of degree 2) in $g$ and $\rho$. As usual in such cases, the standard substitution $u = \frac{g}{\rho}$ gives
\be
\frac{d\rho}{\rho} = \left( \frac{Au}{4u^2 - 1} - u
\right)^{-1} du
\ee
and allows to write the explicit solution:
\be
g \Big[ (A+1) \rho^2 - 4 g^2 \Big]^{A/2} = \text{const}
\label{gABzero}
\ee
The case $A=4$ is precisely the solution of Harvey and Lawson:
\be
M_4 \; = \; \Big\{ \, (g qi\bar q, \rho \bar q) \in \text{Im} \, \mathbb{H} \oplus \mathbb{H}\,, \quad q \in Sp(1) \,, \quad g (4g^2 - 5\rho^2)^2 = \epsilon \; \Big\}
\label{HLsol}
\ee
Topologically, this coassociative 4-manifold is a spin bundle over $S^2$. Therefore, as expected, our coassociative submanifolds defined by the ansatz \eqref{ansatz-y} generalize the Harvey-Lawson construction by replacing $\R^7$ with a more general 7-manifold $X_7 = \R^3 \times \text{TN}_n$.

\subsection{Solving the Abel equation}

Now, let us study solutions to \eqref{ode-final} and \eqref{ode-AB} more systematically. Motivated by the special solutions \eqref{NS5sol} and \eqref{HLsol} we first consider the behavior of $g$ in the limits of small and large $r$.

\subsubsection*{Asymptotic behavior}

\paragraph{Small $r$:}
In the limit $r\to 0$, the equation \eqref{ode-AB} can be approximated by
\beq
{dg \over dr} = {A \, g \over g^2 -2r} ,
\eeq
whose solution is
\beq
r={g^2 \over 2+2A} + c_1 \, g^{-2/A},
\eeq
where $c_1$ is an integration constant determined by the boundary conditions.

For concreteness, let us consider a {\it separatrix} solution that corresponds to the initial condition $g(0)=0$, {\it i.e.} $c_1=0$. We thus get
\beq
g=\sqrt{(2+2A)r}.
\label{g_small_r}
\eeq

\paragraph{Large $r$:}
Assuming that $g$ grows faster than $r^{1/2}$ as $r\to \infty$, an assumption that we shortly verify to be self-consistent, \eqref{ode-AB} becomes:
\beq
{dg \over dr} = {r \over g } .
\eeq
where we set $B=1$ by scaling symmetries \eqref{grscaling}.
The solution to this approximate equation is
\beq
g=\sqrt{r^2 + \zeta} .
\label{g_large_r}
\eeq
If we are interested in the leading behavior at large $r$, the integration constant $\zeta$ can be neglected.

\subsection*{Full solution: the interpolating function}

\begin{figure}[ht]
	\centering
	\includegraphics[width=7.5cm]{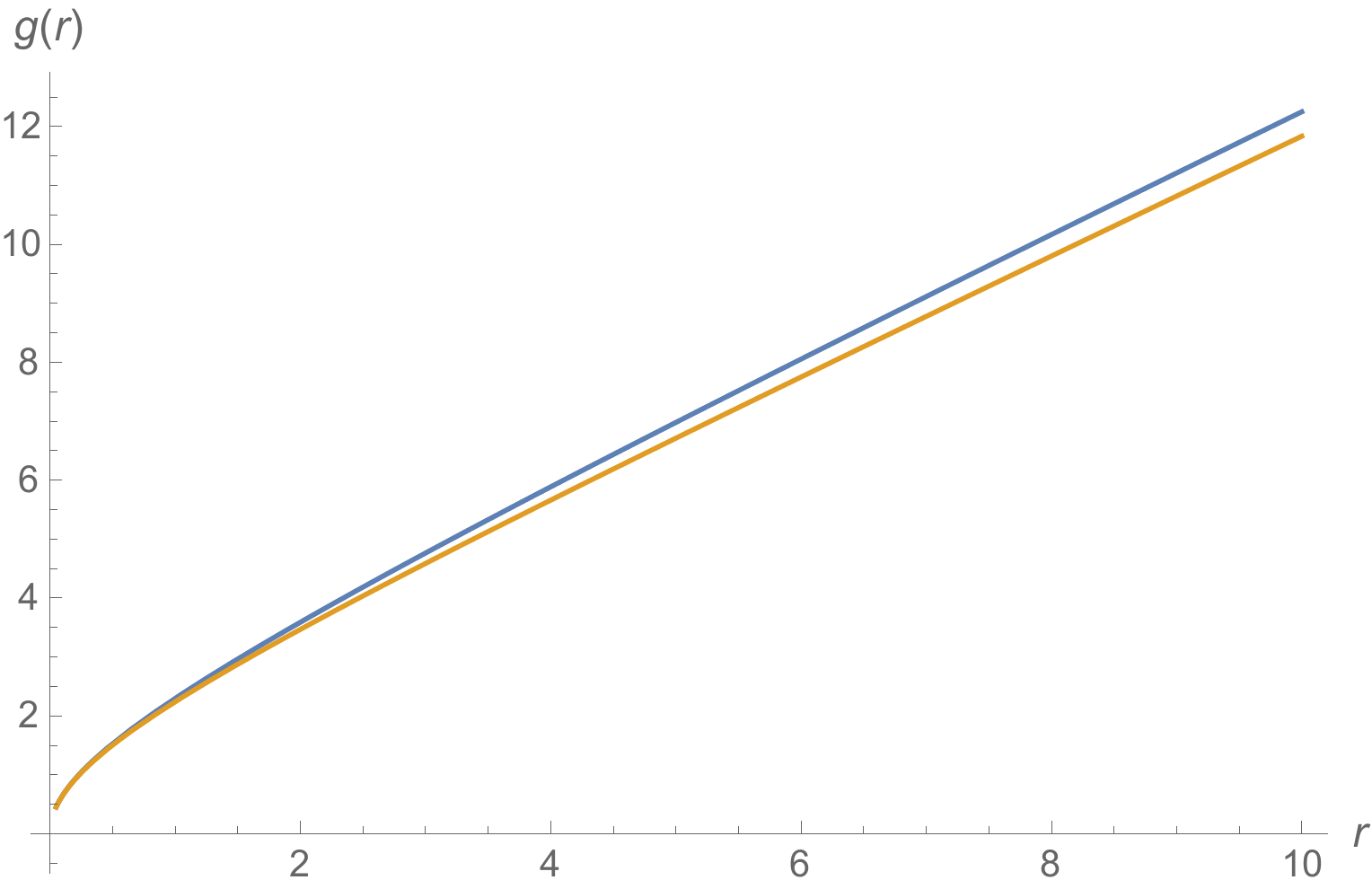}
	\caption{Comparison between the numerical solution for $g$ (blue) and $g_0$ (orange) for $A=1$.} 
	\label{g_vs_g0_A=1}
\end{figure}

Let us introduce the following function 
\beq
g_0=\sqrt{(2+2A)r+r^2} ,
\label{g0fncn}
\eeq
which reproduces the small and large $r$ asymptotic behavior of $g$. Figure~\ref{g_vs_g0_A=1} presents a comparison between the numerical solution for $g$ and $g_0$ for $A=1$. The two functions rapidly become indistinguishable for larger values of $r$. We will later discuss the accuracy of this approximation in further detail and its dependence on $A$.

It is convenient to decompose $g$ into a product as follows
\beq
g=g_0 f ,
\label{g_0xf}
\eeq
where $g_0$ takes care of the asymptotic behavior as $r\to0$ and $\infty$. We refer to $f$ as the {\it interpolating function}. It is clearly a function with the boundary values $f(0)=f(\infty)=1$. Plugging \eqref{g_0xf} into \eqref{ode-AB}, we obtain the following differential equation for $f$
\beq
{df \over dr} = {f \over r} \left[{(A+r) \over (2+2A+r)f^2 - 2} - {(1+A+r) \over (2+2A+r)} \right] .
\label{eq_f}
\eeq

\subsection*{The $A=1$ case}

\begin{figure}[ht]
	\centering
	\includegraphics[width=9cm]{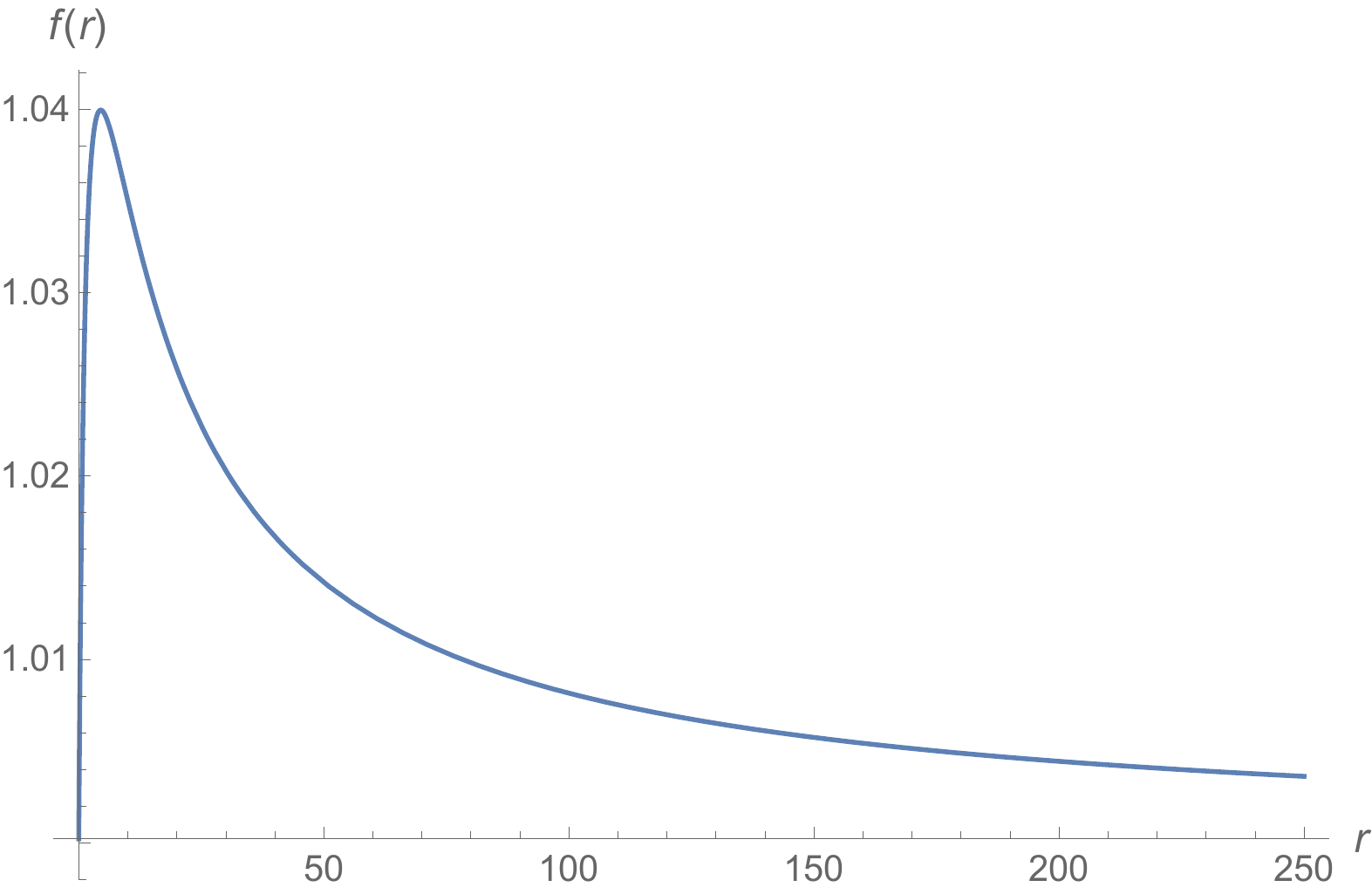}
	\caption{Numerical solution for the interpolating function $f$ with $A=1$.}
	\label{f_numerical_A=1}
\end{figure}

Let us first consider the case of $A=1$. Figure~\ref{f_numerical_A=1} shows the function $f$, obtained by numerically solving \eqref{eq_f}. It is interesting to note that the maximum value of $f$ is around $1.04$, which means that in this case $g_0$ is already a reasonably good approximation to $g$, differing from it by at most $4\%$.

\begin{figure}[ht]
	\centering
	\includegraphics[width=9cm]{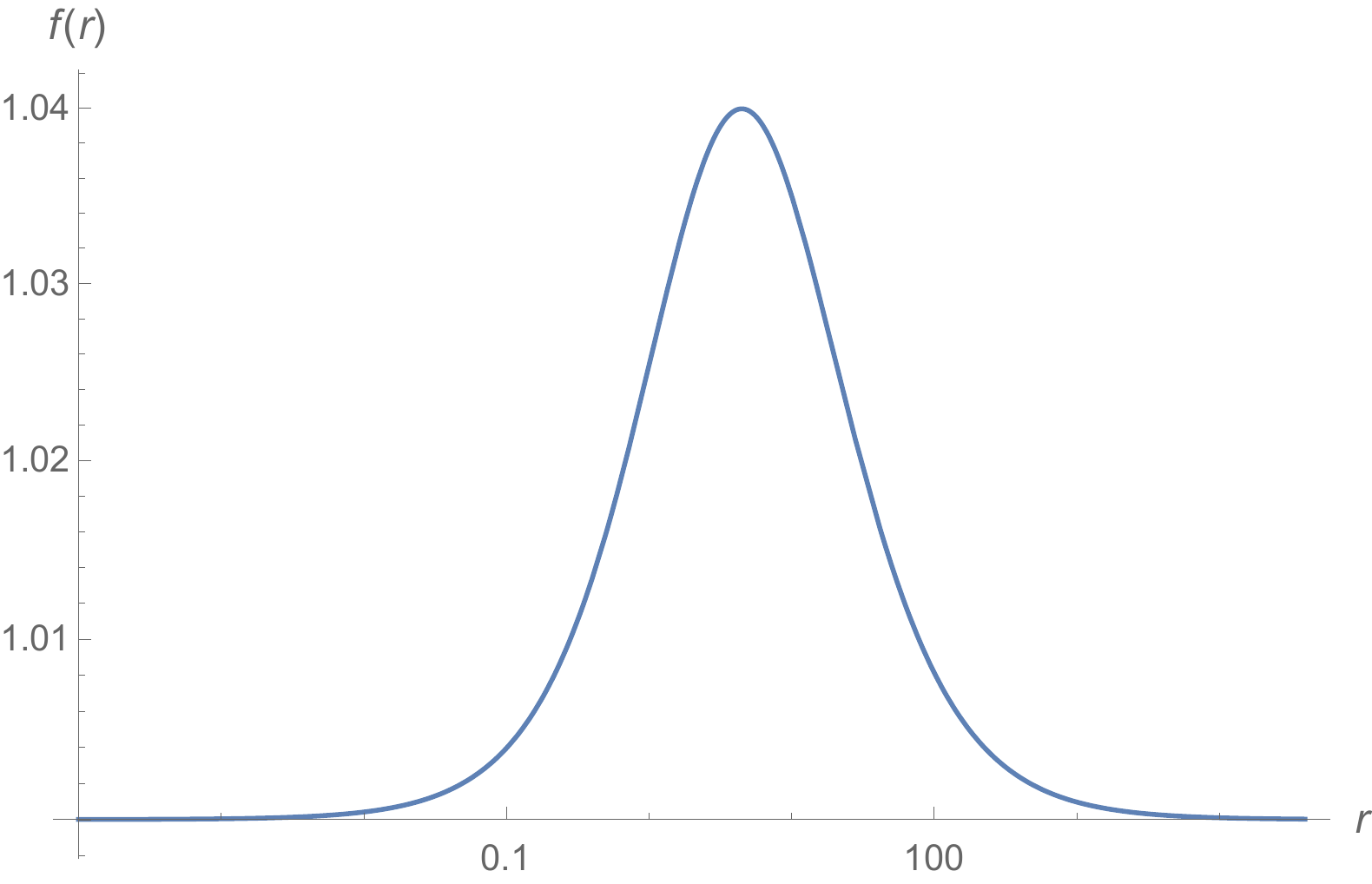}
	\caption{Log-log plot of the numerical solution for the interpolating function $f$ with $A=1$.}
	\label{f_numerical_A=1_loglog}
\end{figure}

This function becomes remarkably simple and suggestive when plotted in log-log scale, as shown in Figure~\ref{f_numerical_A=1_loglog}.

\paragraph{An approximate solution.}
Motivated by the simplicity of Figure~\ref{f_numerical_A=1_loglog}, we propose an analytical approximation $f_{app}$ to $f$. This is given by the 4-parameter function
\beq
f_{app}=(1 - \alpha) + \alpha \, e^{C \, e^{-{(\log(r) - \log(r_0))^2\over 2 \Delta^2}}} .
\label{f_app}
\eeq
Let us briefly motivate this expression. First of all, at $\alpha=1$, it gives a Gaussian function in log-log scale, which is a natural ansatz in view of Figure~\ref{f_numerical_A=1_loglog}. To achieve a better fit, we introduced an additional parameter $\alpha$. From the point of view of the second term, in log-log scale it would correspond to a vertical shift by $\log \alpha$. The first term guarantees that $f_{app}(0)=f_{app}(\infty)=1$.

\paragraph{Parameter fit.}

\begin{figure}[ht]
	\centering
	\includegraphics[width=9cm]{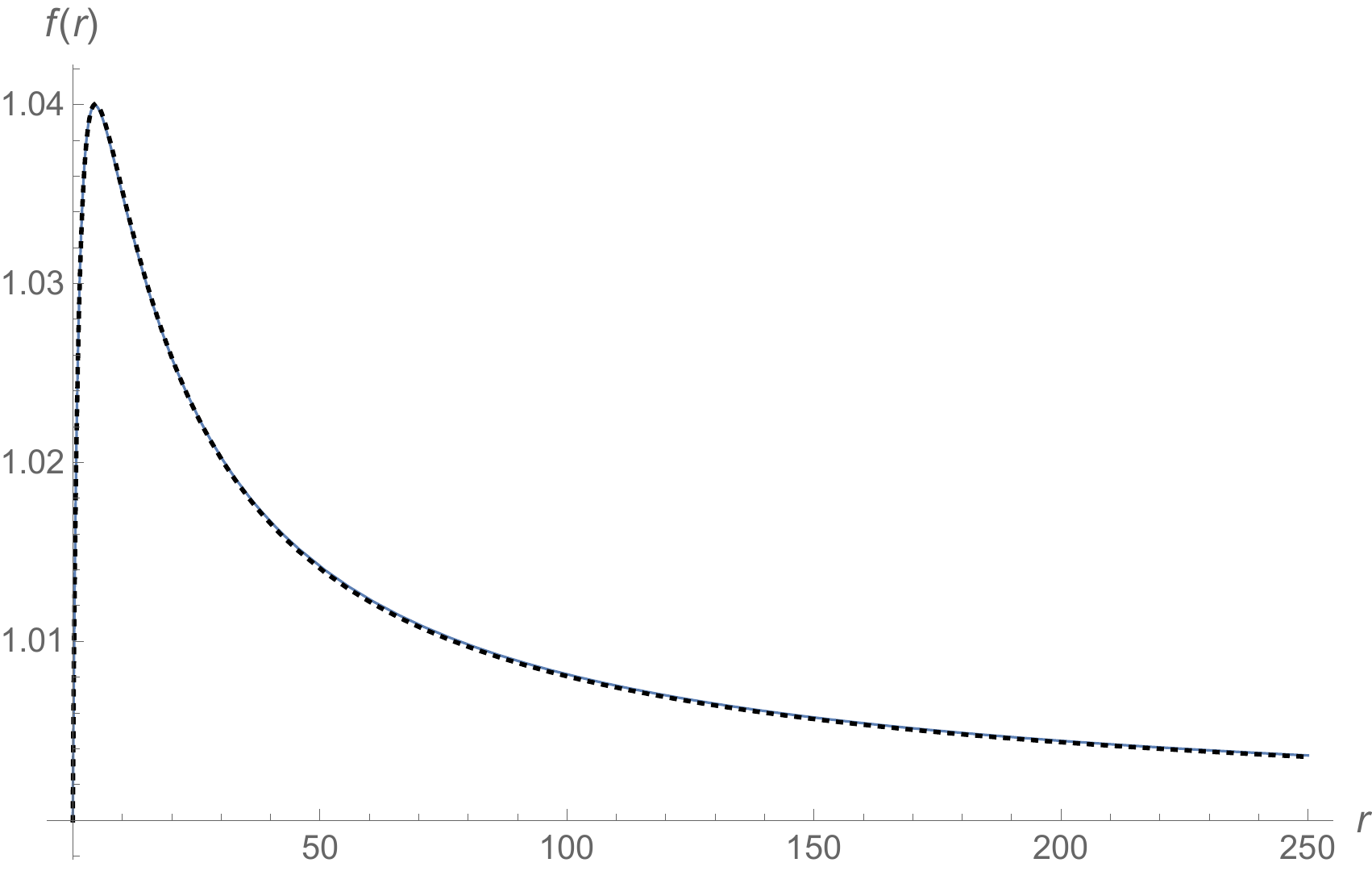}
	\caption{Comparison between the numerical solution for $f$ (blue) and the fitted $f_{app}$ (dotted) for $A=1$.}
	\label{f_numerical_fit_A=1}
\end{figure}

\begin{figure}[ht]
	\centering
	\includegraphics[width=9cm]{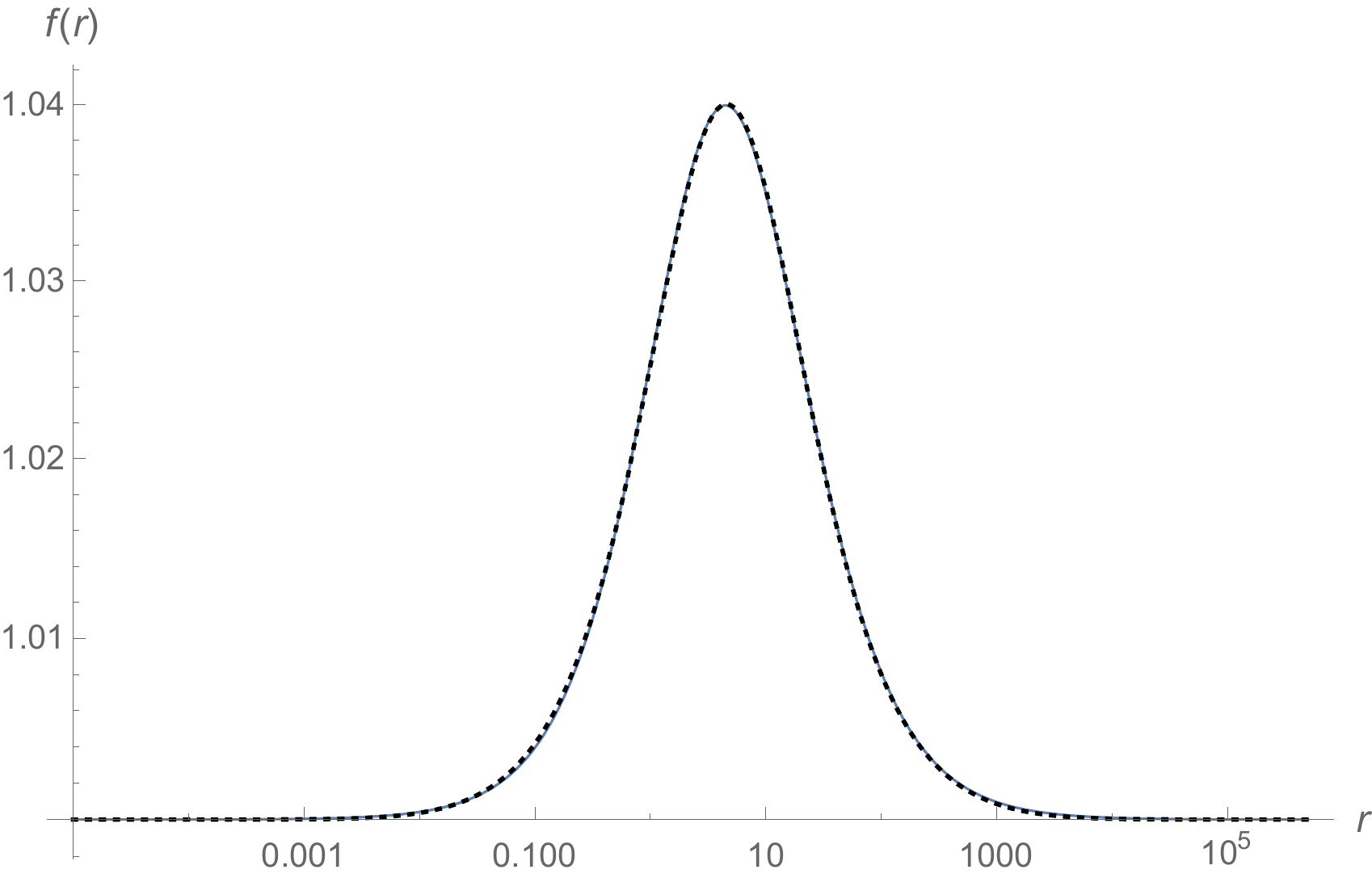}
	\caption{Log-log plot of the comparison between the numerical solution for $f$ (blue) and the fitted $f_{app}$ (dotted) at $A=1$.}
	\label{f_numerical_fit_A=1_loglog}
\end{figure}

Figures~\ref{f_numerical_fit_A=1} and \ref{f_numerical_fit_A=1_loglog} compare $f$ to the best fit of the numerical solution by $f_{app}$. The fit was obtained by sampling the numerical solution at $500$ points equally spaced in $\log r$, between $r=10^{-5}$ and $r=10^5$. Increasing the number of sample points does not noticeably modify the fit. The fitted parameters for this case are: $\alpha=0.00726475$, $C=1.87302$, $r_0=4.59284$, and $\Delta=2.26916$.

The function $\eqref{f_app}$ provides an excellent, albeit not perfect, fit to $f$. Indeed, it is possible to verify that there is no choice of parameters such that $f_{app}$ is a solution of \eqref{eq_f}. It would be interesting to investigate whether a small modification of it yields an exact solution.

\begin{figure}[ht]
	\centering
	\includegraphics[width=9cm]{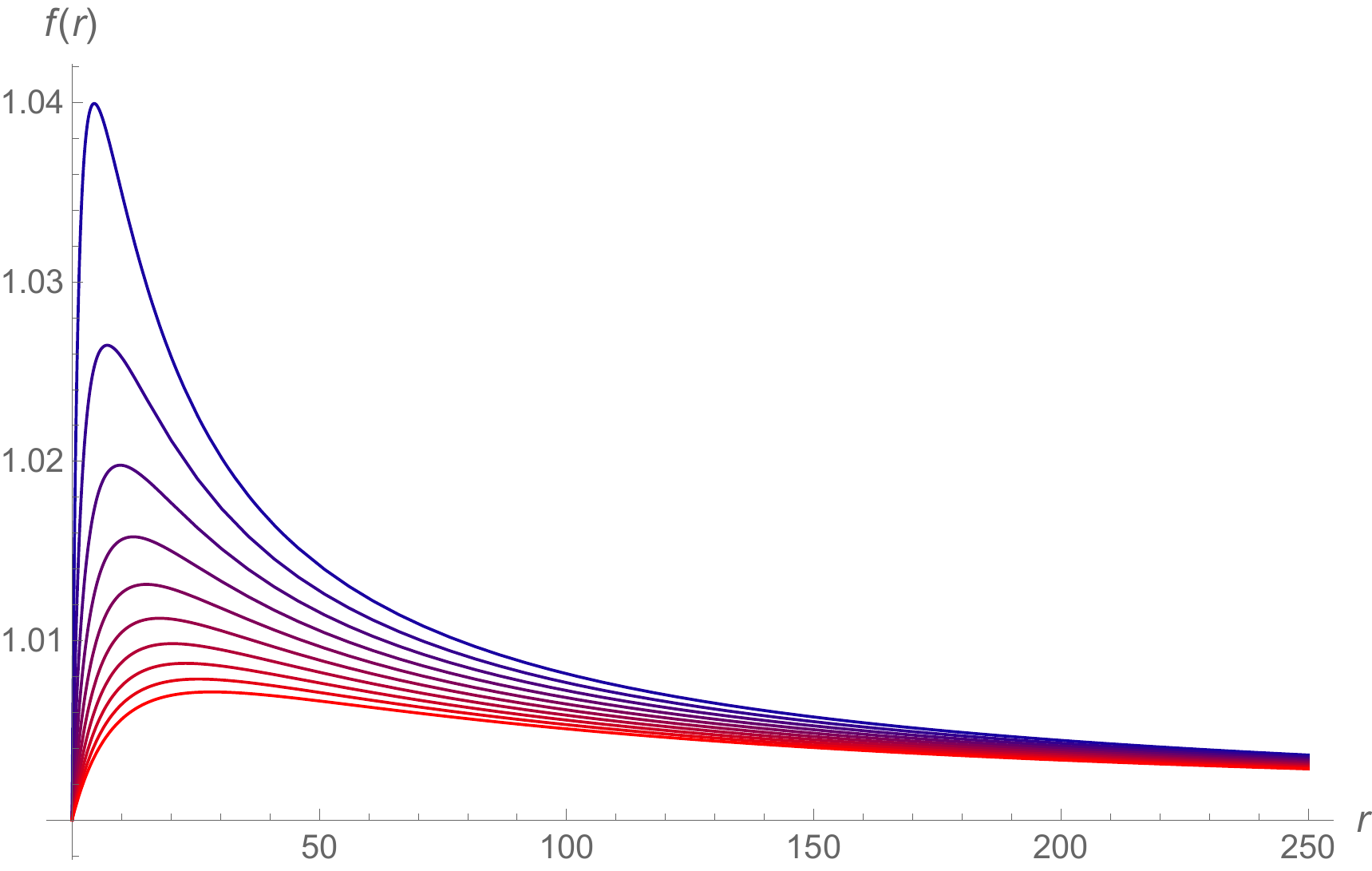}
	\caption{Numerical solution for $f$ for $A=1$ (blue) to $A=10$ (red).}
	\label{f_numerical_A=1to10}
\end{figure}

\subsection*{General $A$}

\begin{figure}[ht]
	\centering
	\includegraphics[width=9cm]{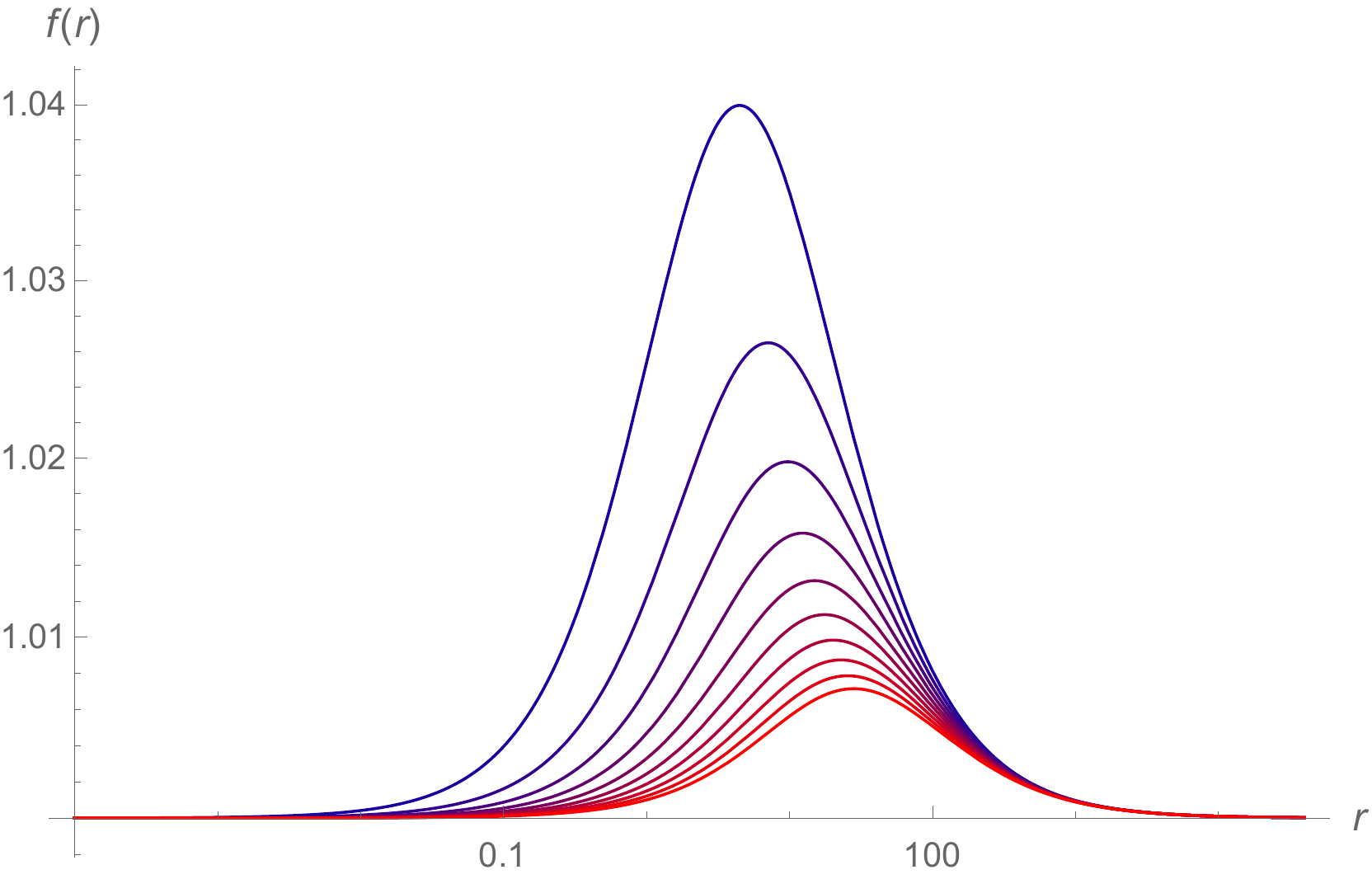}
	\caption{Log-log plot of the numerical solution for $f$ for $A=1$ (blue) to $A=10$ (red).}
	\label{f_numerical_A=1to10_loglog}
\end{figure}

We now consider the case of general $A$, including the value $A=4$ of interest in \eqref{ode-final}. Since the analysis is identical to the one for $A=1$, our presentation will be more succinct. Figure~\ref{f_numerical_A=1to10} shows $f$ for various values of $A$ ranging from $1$ to $10$, obtained by numerically solving \eqref{eq_f}.
Once again, $f$ becomes extremely simple in log-log scale, as shown in Figure~\ref{f_numerical_A=1to10_loglog}.

\begin{figure}[ht]
	\centering
	\includegraphics[width=9cm]{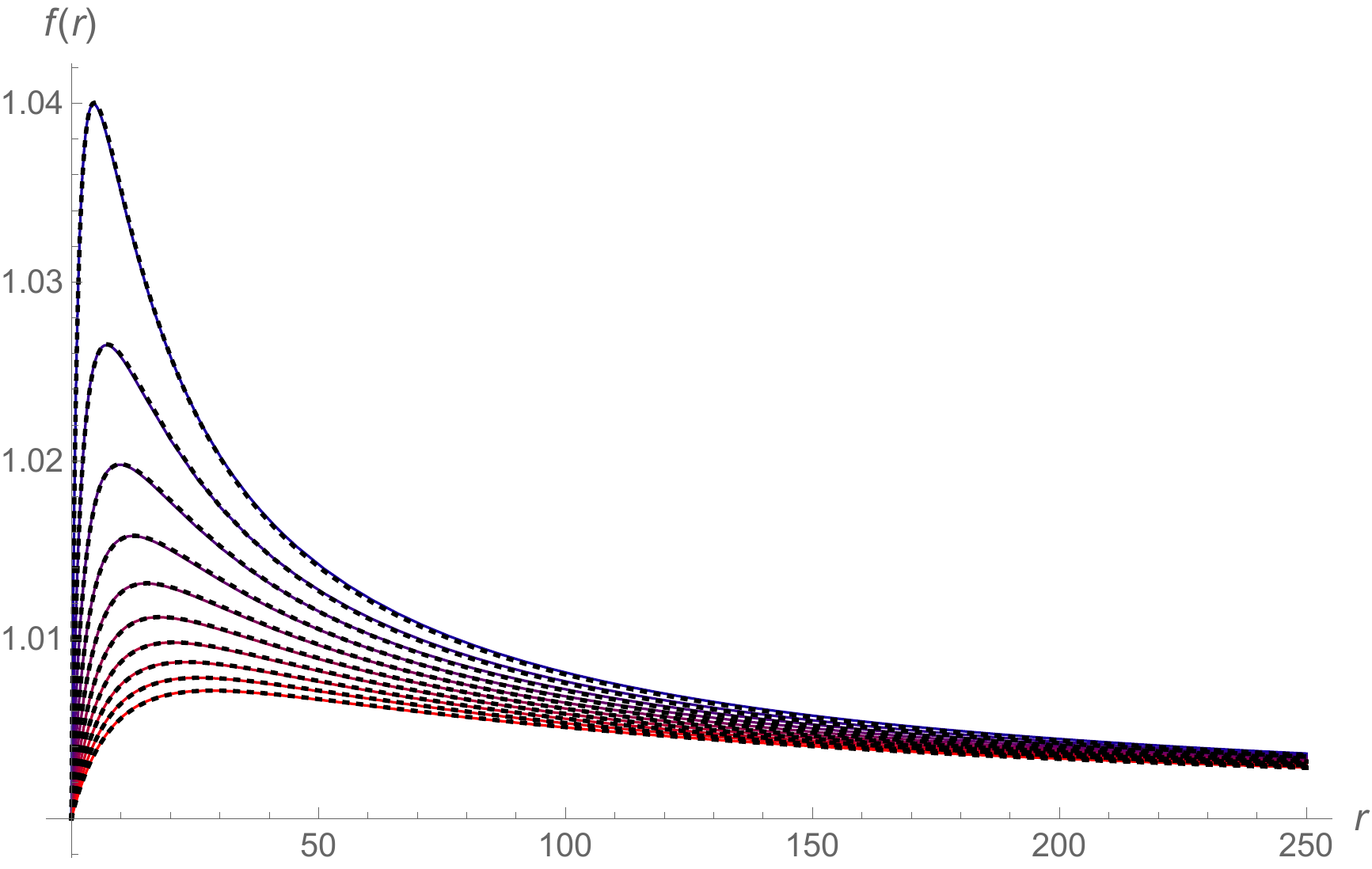}
	\caption{Comparison between $f$, for $A=1$ (blue) to $A=10$ (red), to the fitted $f_{app}$ (dotted).}
	\label{f_numerical_fit_A=1to10}
\end{figure}

\begin{figure}[ht]
	\centering
	\includegraphics[width=9cm]{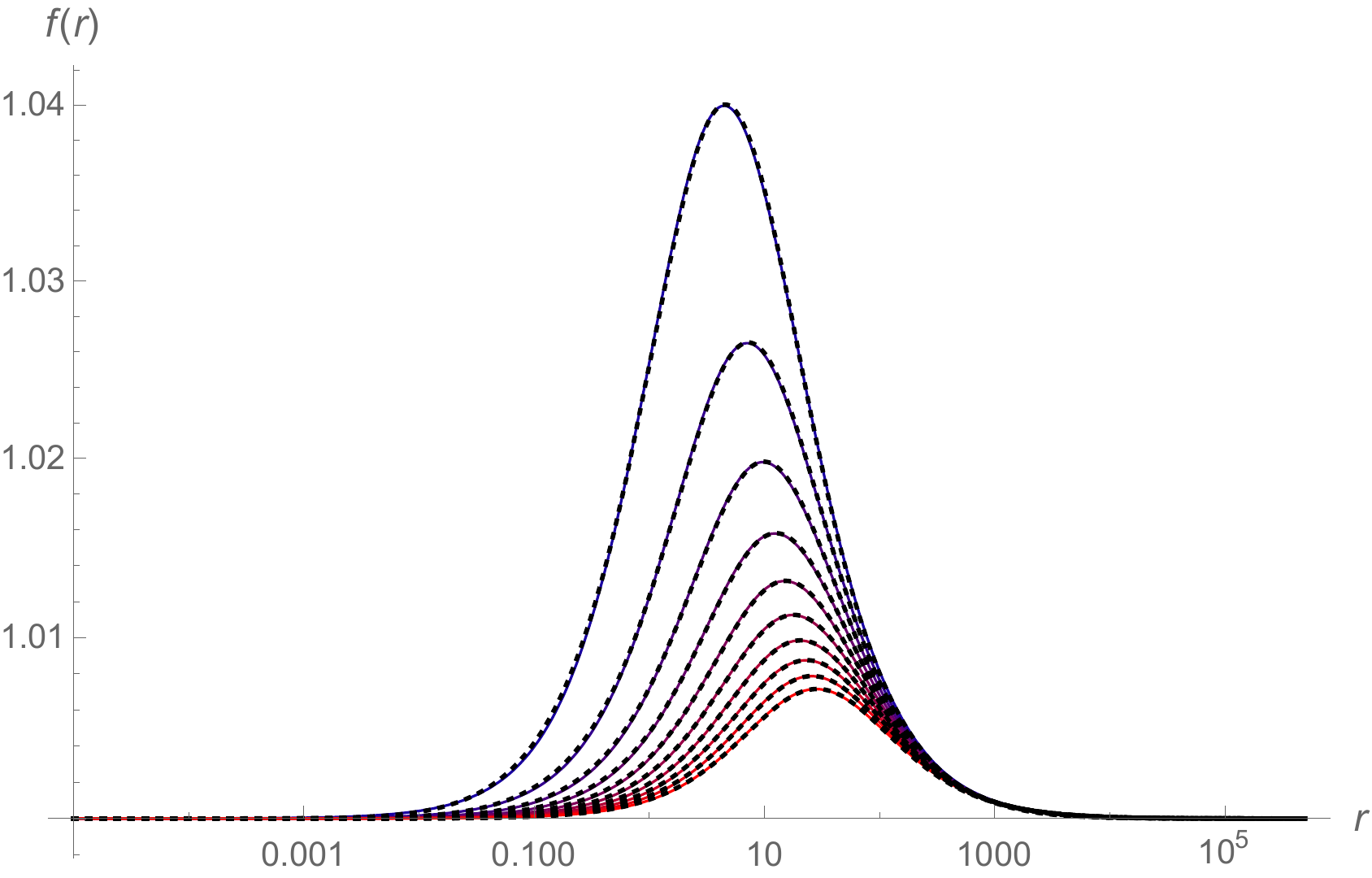}
	\caption{Log-log plot of the comparison between $f$, for $A=1$ (blue) to $A=10$ (red), to the fitted $f_{app}$ (dotted).}
	\label{f_numerical_fit_A=1to10_loglog}
\end{figure}

\paragraph{Parameter fit.}

As in the case $A=1$, $f_{app}$ provides an excellent approximation to $f$ for all $A$. Figures \ref{f_numerical_fit_A=1to10} and \ref{f_numerical_fit_A=1to10_loglog} compare $f$ for values of $A$ in the range $1, \ldots, 10$ to their fits by $f_{app}$.
The parameters have been fitted using the same method as before. Below we present them for reference.

\beq
\begin{array}{|c|c|c|c|c|}
	\hline
	A & \alpha & C & r_0 & \Delta \\ \hline
	1 & 0.00726475 & 1.87302 & 4.59284 & 2.26916 \\
	2 & 0.00412429 & 2.00496 & 7.28069 & 2.28823 \\
	3 & 0.00286021 & 2.06973 & 9.97222 & 2.29881 \\
	4 & 0.00218574 & 2.1077 & 12.665 & 2.30526 \\
	5 & 0.00176766 & 2.13258 & 15.3582 & 2.30959 \\ 
	6 & 0.00148362 & 2.15003 & 18.0519 & 2.31264 \\
	7 & 0.00127792 & 2.16309 & 20.7456 & 2.31497 \\
	8 & 0.00112233 & 2.17309 & 23.4394 & 2.31675 \\ 
	9 & 0.00100046 & 2.18105 & 26.1333 & 2.31818 \\
	\ \ 10 \ \ \ & \ \ \ 0.000902438 \ \ \ \ & \ \ \ 2.18753 \ \ \ \ & \ \ \ 28.8273 \ \ \ \ & \ \ \ 2.31934 \ \ \ \  \\
	\hline
\end{array}
\eeq

\bigskip

\subsection*{The full function}

We are now ready to put together the asymptotic function $g_0$ with the analytic approximation to the interpolating function $f_{app}$ and compare it to $g$.

\paragraph{Comparison to $g_0$.}

\begin{figure}[h]
	\centering
	\begin{minipage}{0.45\textwidth}
		\centering
		\includegraphics[width=0.8\textwidth]{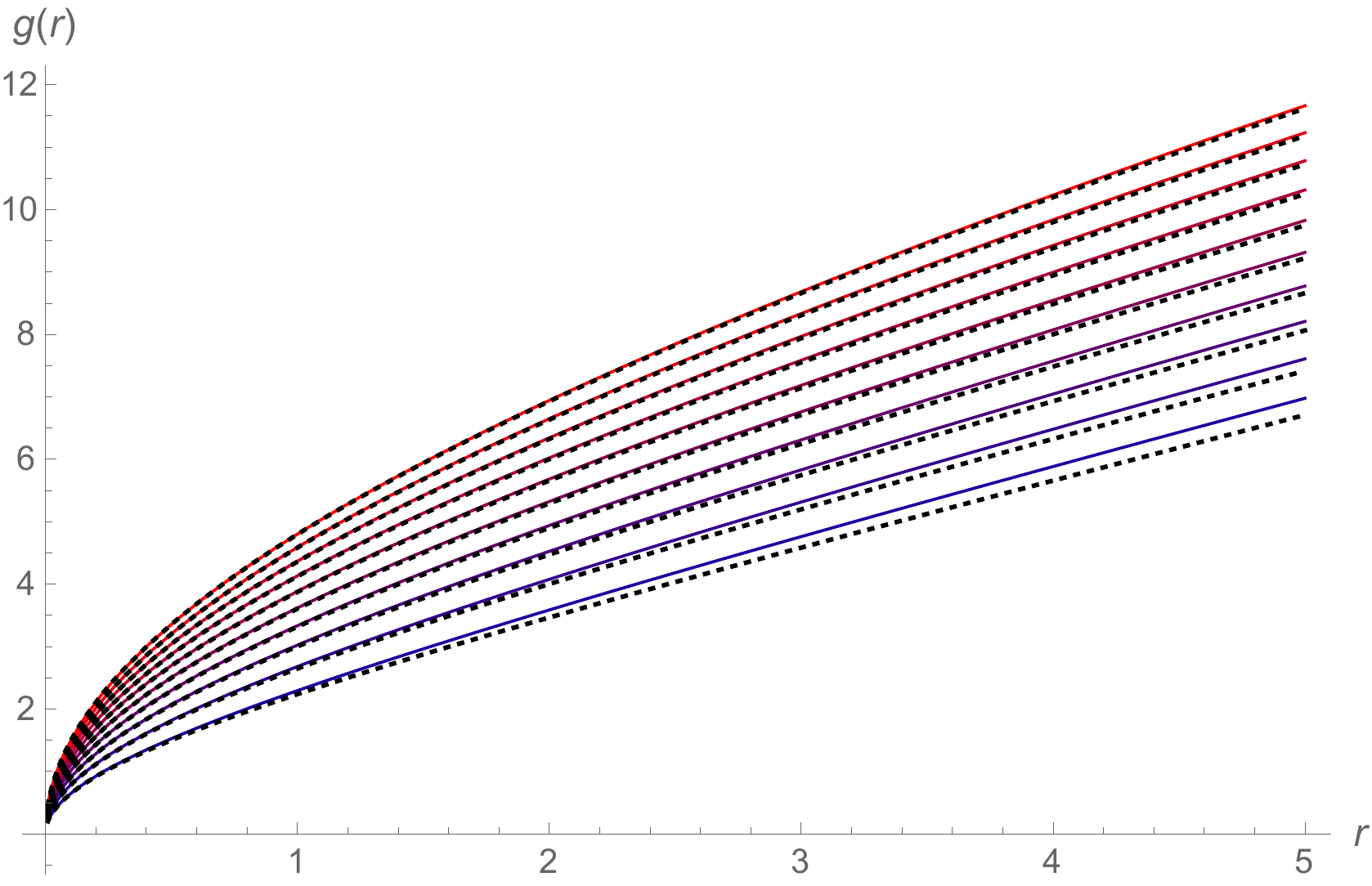}
		\notag
		\label{fig:sub1}
	\end{minipage}
	\qquad
	\begin{minipage}{0.45\textwidth}
		\centering
		\includegraphics[width=0.8\textwidth]{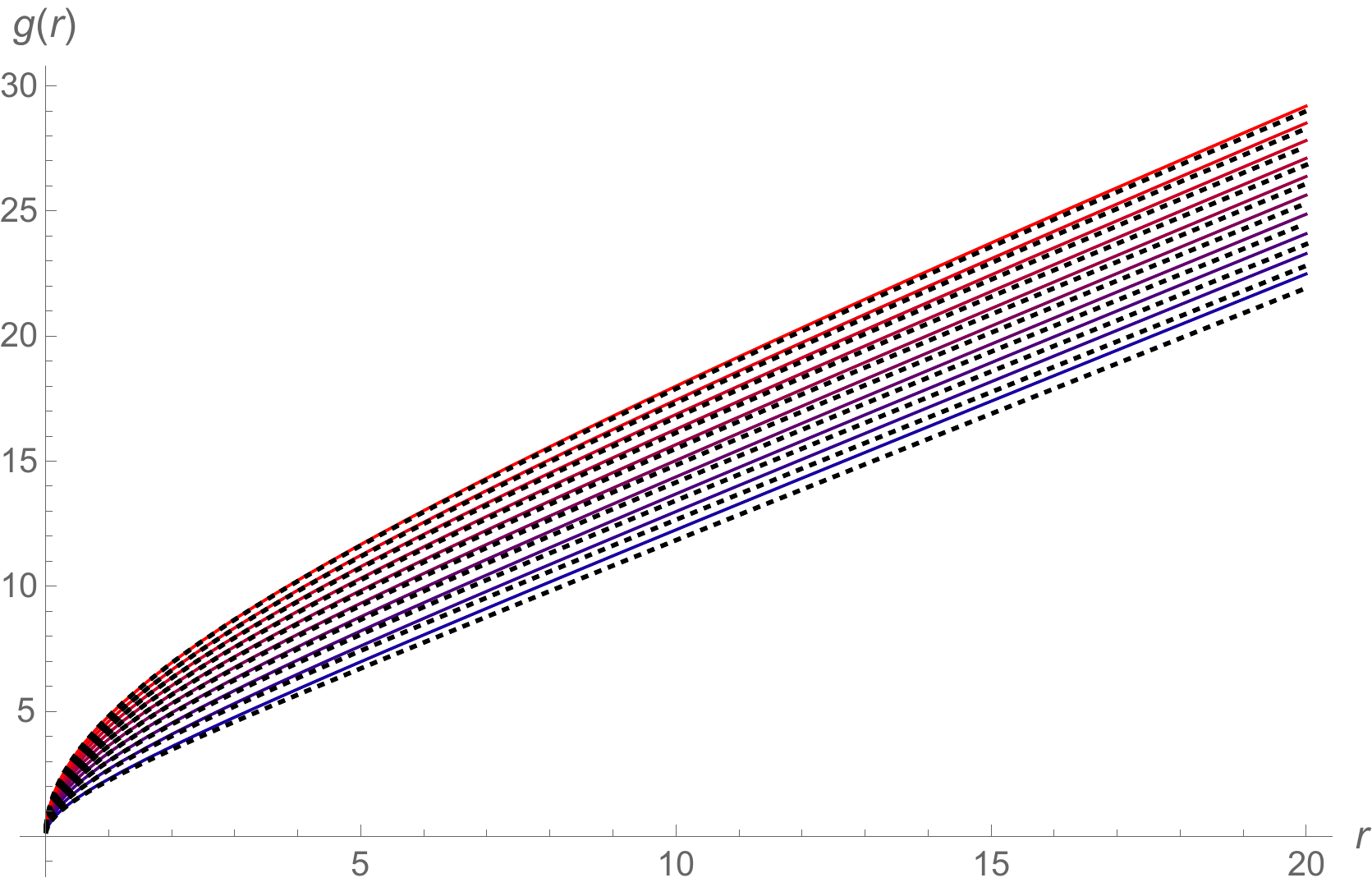}
		\notag
		\label{fig:sub2}
	\end{minipage}
\caption{Comparison between $g$, for $A=1$ (blue) to $A=10$ (red), to $g_0$ (dotted). We consider the ranges: a) $0\leq r \leq 5$ and b) $0\leq r \leq 20$.}
\label{g_vs_g0}
\end{figure}

In order to appreciate how things improve by introducing $f_{app}$, it is convenient to first compare $g$ to $g_0$, as shown in Figure~\ref{g_vs_g0}. We show two ranges of small $r$, that is where the deviations are most noticeable. We see that $g_0$ becomes a better approximation to $g$ as $A$ increases. We will revisit this observation shortly.


\paragraph{Comparison to $g_0 f_{app}$.}

\begin{figure}[h]
	\centering
	\begin{minipage}{0.45\textwidth}
		\centering
		\includegraphics[width=.95\linewidth]{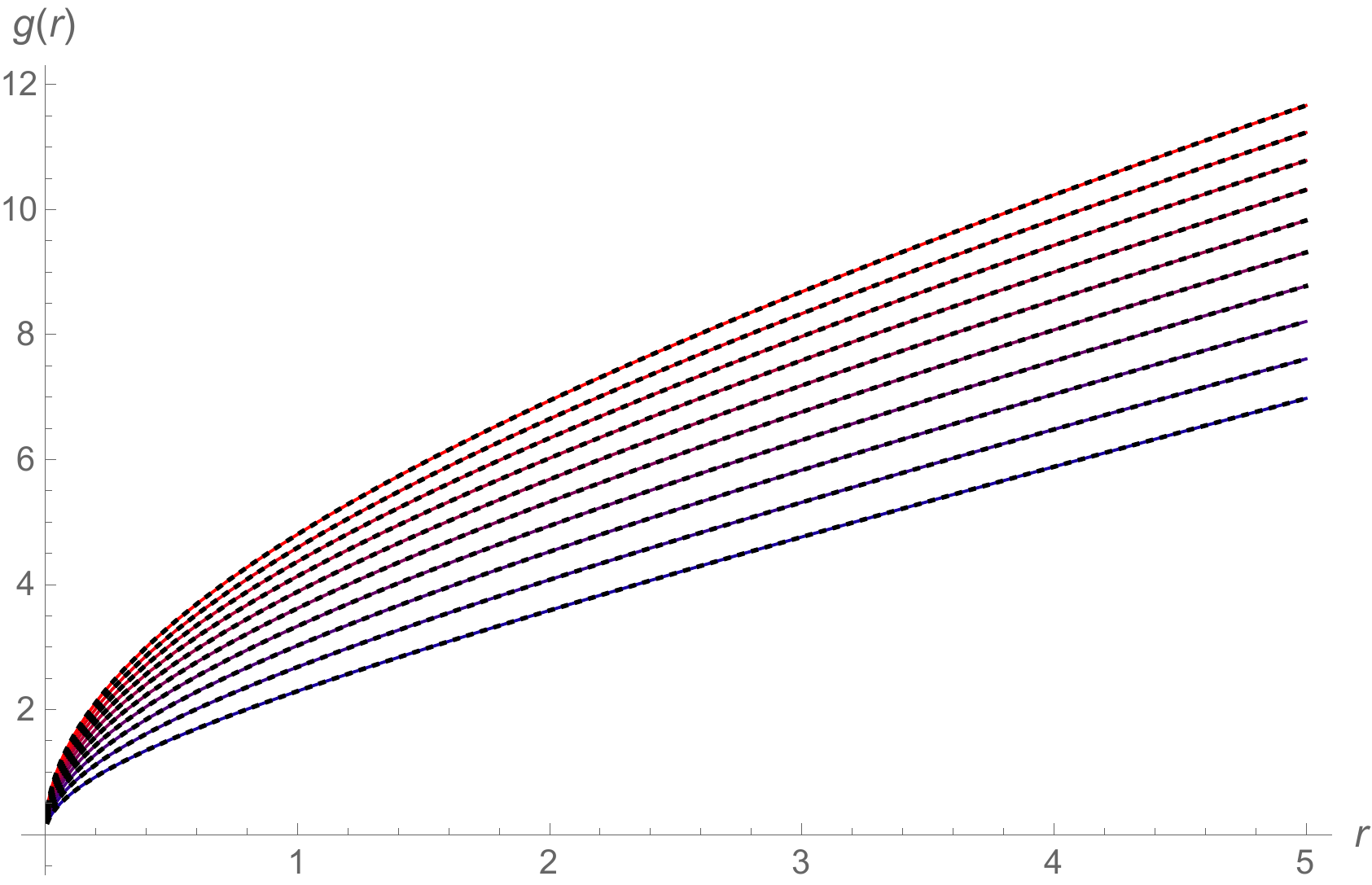}
		\notag
		\label{fig:sub1}
	\end{minipage}
	\qquad
	\begin{minipage}{0.45\textwidth}
		\centering
		\includegraphics[width=.95\linewidth]{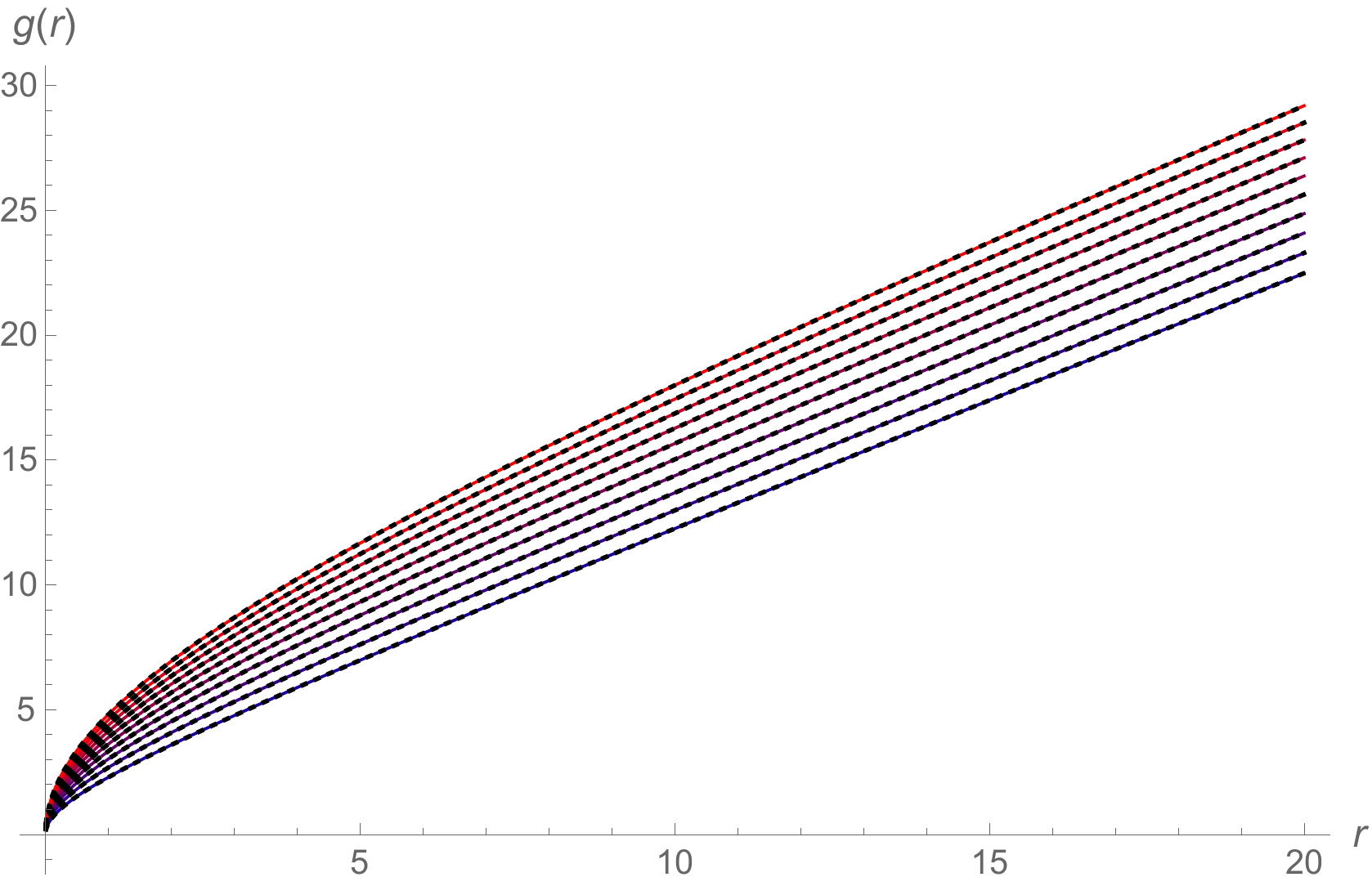}
		\notag
		\label{fig:sub2}
	\end{minipage}
	\caption{Comparison between $g$, for $A=1$ (blue) to $A=10$ (red), to $g_0 f_{app}$ (dotted). We consider the ranges: a) $0\leq r \leq 5$ and b) $0\leq r \leq 20$.}
	\label{g_vs_g0fapp}
\end{figure}

Figure~\ref{g_vs_g0fapp} compares $g$ to the analytic approximation given by $g_0 f_{app}$ for $A=1, \ldots, 10$. The functions become indistinguishable to a naked eye.

\subsection*{The $A\to \infty$ limit.}

Figures \ref{f_numerical_A=1to10} and \ref{f_numerical_A=1to10_loglog} show that the maximum value of $f$ decreases as $A$ grows. In other words, $g_0$ becomes a better approximation to $g$ as $A$ is increased. The maximum discrepancy between the two functions goes from $4\%$ for $A=1$ to $0.7\%$ for $A=10$. For $A=40$ this number reduces to $0.2\%$. 

We can discuss this behavior in terms of $f_{app}$, since it gives a good approximation to $f$. The maximum of $f_{app}$ is $f_{app}(r_0)=1+\alpha(e^B-1)$. Computing the fits up to $A=40$, we observe that $B$ seems to converge to a finite value while $\alpha$ appears to decrease to zero. This leads us to conjecture that $f_{app}\to 1$ as $A\to \infty$ or, equivalently, that $g_0$ becomes the exact solution in the $A\to \infty$ limit. Notice that this is not just the small $r$ solution \eqref{g_small_r}, since for sufficiently large $r$ the behavior of $g$ is controlled by $B$.

\begin{figure}[ht]
	\centering
	\includegraphics[width=9cm]{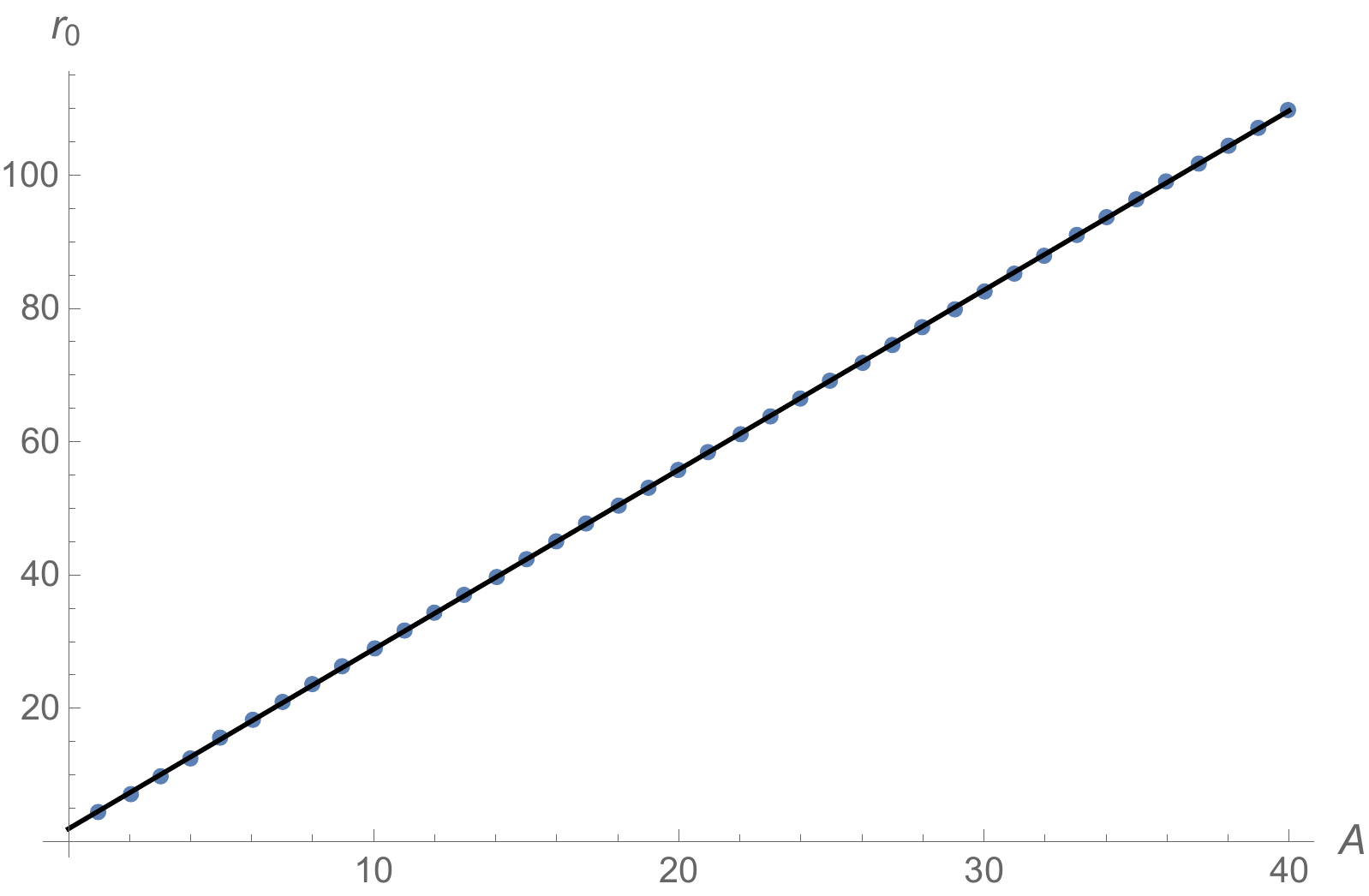}
	\caption{Value of $r_0$ as a function of $A$ up to $40$ and its linear fit.}
	\label{r0_vs_A}
\end{figure}

We can offer more details on how this limit is approached. The interpolating function $f$ accounts for the transition between the small and large $r$ regimes of $g$. The value of $r_0$ is a natural indicator of where this transition occurs. Interestingly, $r_0$ depends linearly on $A$, as shown in Figure~\ref{r0_vs_A}. As $A$ grows, not only $f_{app}$ approaches 1, signaling that $g_0$ becomes a better approximation to $g$, but also $r_0$ grows, indicating that the small $r$ approximation is valid up to larger values of $r$, as expected.

\subsection*{$L^2$ norms}

For the Harvey-Lawson solution \eqref{HLsol}, it was shown in \cite{Gukov:2002zg} that the deformation mode $\epsilon$ is not $L^2$ normalizable:
\be
\int_{M_4} d^4 x \sqrt{\det (g_{\mu\nu} + \delta g_{\mu\nu})} - \int_{M_4} d^4 x \sqrt{\det (g_{\mu\nu})} \; \to \; \infty
\ee
For $B=0$ and more general (integer) values of $A$, from \eqref{gABzero} we find
\be
\frac{\delta g}{\delta \epsilon} \; = \; 
\frac{1}{(A+1)(\rho^2 - 4g^2)} \left( \frac{g}{\epsilon} \right)^{\frac{A-2}{A}}
\; \sim \; \rho^{-\frac{A+2}{A}}
\qquad \text{as} \quad \rho \to \infty
\label{L2normA}
\ee
This exhibits a faster than $\rho^{-2}$ decay when $A<2$, which is the condition for the deformation mode $\epsilon$ to be $L^2$ normalizable. If $A>2$, the $\epsilon$-mode is not $L^2$ normalizable and usually this means that it should be interpreted as a parameter (coupling constant) of the IR theory, rather than a dynamical field with finite kinetic term.

This is not the full story, however, in our present context. First, our IR theory is a two-dimensional $\CN=(0,2)$ theory that ``lives'' on the $\R^2$ part of the fivebrane world-volume~\eqref{M5branes}. As is well known, in two space-time dimensions, long range quantum fluctuations do not allow fixing vacuum expectation values, and so all moduli tend to be dynamical. A natural question, then, is whether a singularity $\epsilon=0$ occurs at finite distance in the moduli space metric, see {\it e.g.} \cite{Gukov:1999ya} for a close cousin of our problem where ``compactification'' on a non-compact 4-manifold is replaced by a ``compactification'' on a non-compact Calabi-Yau 4-fold.

Another obvious modification of the estimate \eqref{L2normA} is due to the fact that it was deduced under the assumption $B=0$. Indeed, as we noted several times earlier, the large-$r$ behavior of the solution is controlled entirely by $B$, not $A$. Thanks to the detailed analysis in this section, however, it is easy to see that incorporating $B \ne 0$ still leads to the same conclusion as \eqref{L2normA}. Namely, from \eqref{g0fncn}--\eqref{g_0xf} and \eqref{f_app} we quickly find the large-$r$ behavior of $g \sim r$ and $\frac{\delta g}{\delta \zeta} \sim r^{-1}$, which means that the deformation mode is not $L^2$ normalizable.

\subsection{Reversing the degree}

The differential equation \eqref{ode-final} and its solutions described above correspond to D4-NS5 brane models with non-compact D4-branes of the form \eqref{diskcompl}. Moreover, as explained in the end of section \ref{sec:anomaly}, brane models with compactly supported D4-branes, as in \eqref{D4-disk}, differ by the orientation reversal on the fibers of $X_7$ or, equivalently, $n \to - n$.

In our parametrization of the coassociative submanifolds \eqref{ansatz-y}, this corresponds to simultaneously changing the sign of $g$ and $r$. Therefore, the corresponding version of the ODE \eqref{ode-final} reads ($n>0$)
\be
\frac{dg}{dr} 
= \frac{(\textcolor{red}{-}n+ |n| + Br )g}{g^2\textcolor{red}{+}nr} 
= \frac{Br g}{g^2\textcolor{red}{+}nr} \,.
\label{odex-flipped}
\ee
The term $|n|$ is not affected by the sign flip, cancels against $(-n)$, and leads to the main qualitative difference between \eqref{ode-final} and \eqref{odex-flipped}. The $|n|$ term is needed to maintain the components of Taub-NUT metric positive. One important consequence is that, although the resulting configuration is free of the winding number anomaly, it breaks supersymmetry and hence is likely to be unstable. 

\begin{figure}[h!]
	\centering
	\includegraphics[height=6cm]{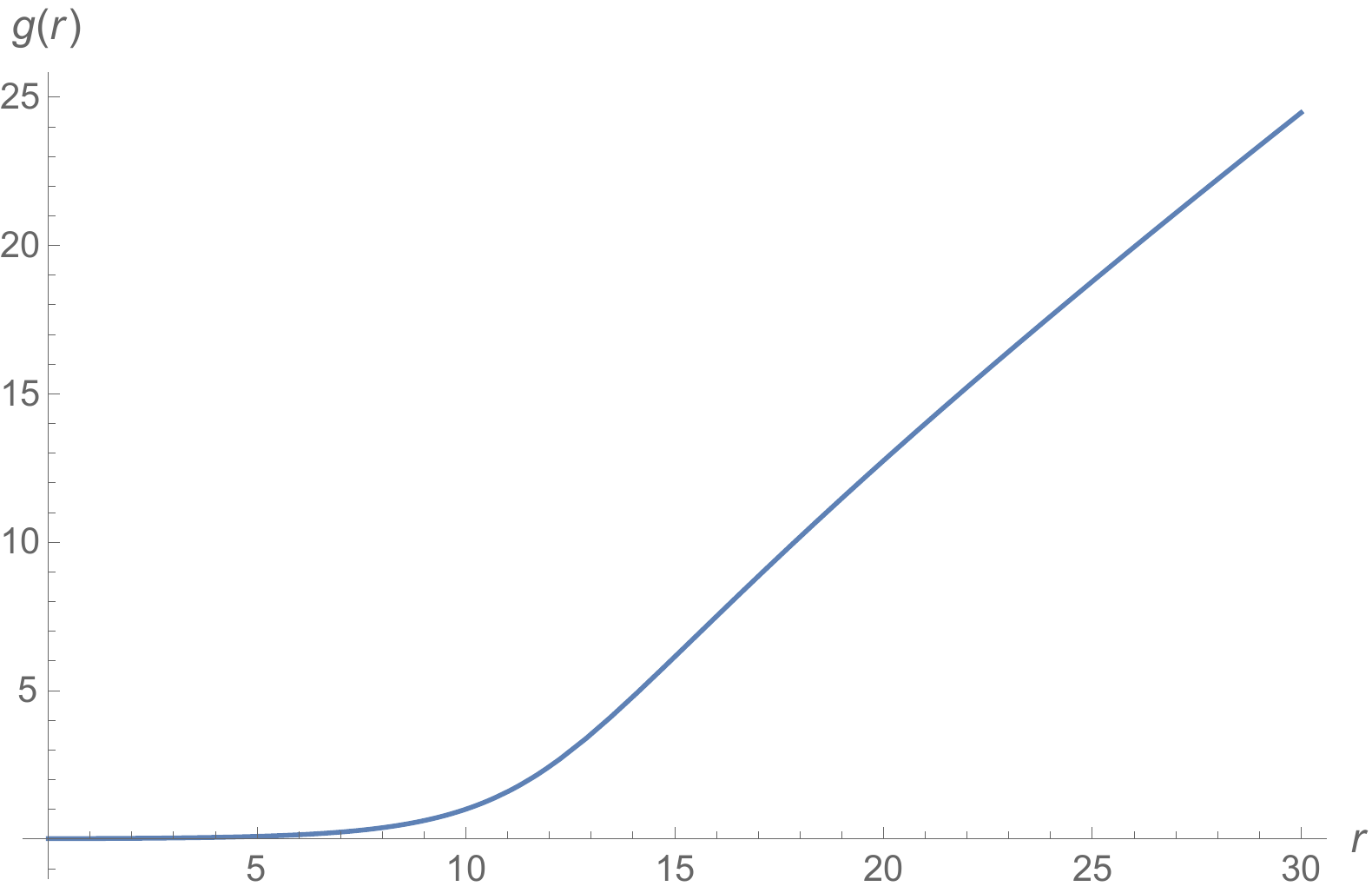}
	\caption{A numerical solution to the differential equation \eqref{odex-flipped}.}
	\label{fig:flip-sol}
\end{figure}

In Figure~\ref{fig:flip-sol}, we plot a numerical solution to the differential equation \eqref{odex-flipped} with $B=1$. Note, just like $SO(3)$-rotation \eqref{D4NS5b}--\eqref{ansatz-y} of the graph of function $g(r)$ discussed earlier in this section produces coassociative submanifolds that correspond to NS5-D4 brane systems with non-compact D4-branes on ``disk complements'' \eqref{diskcompl}, a similar $SO(3)$-rotation of the graph of function $g(r)$ presented in Figure~\ref{fig:flip-sol} clearly has only one asymptotic region $\cong \R_+ \times S^3 / Z_n$ that corresponds to M-theory lift of NS5-branes and, therefore, D4-branes in this brane model are compactly supported on Lagrangian disks~\eqref{D4-disk}.

\section{Phase transitions and $Spin(7)$ manifolds}
\label{sec:physics}

In the previous section, we constructed new coassociative submanifolds $M_4 \subset X_7 = \Lambda^{2,+} (\text{TN}_n)$ of topology $L(n,1) \times \R$, or $\text{TN}_n$, or $\CO (n) \to \cp^1$, or any combinations thereof. These are particular examples of a more general construction \eqref{generalM4M3}, which can be carried out in detail following the steps of section~\ref{sec:ODE}.

As explained in the Introduction, every such pair $(M_4,X_7)$ determines a manifold $X_8$ of $Spin(7)$ holonomy that, in physics, arises from an (auxiliary) problem of D6-branes supported on $\R^3 \times M_4$. The topology of $X_8$ is determined by the topology of $M_4$ and $X_7$. Specifically, $X_8$ is the total space of a circle bundle over $X_7$ with a codimension-4 singular locus $M_4 \subset X_7 \cong X_8 / S^1$. (Note, codimension of the singular locus is always even, and must be equal to 4 in order for the fibration to be smooth.) In our class of examples, this means that the space of $Spin(7)$ holonomy metrics consists of several branches (or phases).

Indeed, consider, for concreteness, the case of $n=1$. Then, one choice of the coassociative submanifold $M_4$ is a disjoint union of a Taub-NUT space $\text{TN}_1$ and $\CO (+1) \to \cp^1$:
\be
\text{Phase A:} \qquad
M_4 \quad = \quad
\begin{matrix}
\CO(+1) \\
\downarrow \\
\cp^1
\end{matrix}
\quad \bigsqcup \quad \text{Taub-NUT}
\label{M4A}
\ee
In this phase, the corresponding $S^1$-fibration has topology\footnote{Sometimes we write non-trivial vector bundles simply as products (or, more precisely, our use of ``$\cong$'' means homotopy eqivalence). For example, $X_8^{(A)}$ here is the total space of the universal quotient bundle over $\cp^2$.} $X_8^{(A)} \cong \R^4 \times \cp^2$ and depends on one real deformation parameter that we can choose to be $\text{Vol} (\cp^2)$. As $\text{Vol} (\cp^2) \to 0$, the $Spin(7)$ holonomy metric on $X_8^{(A)}$ develops an isolated conical singularity which, aside from $X_8^{(A)}$, admits another resolution (``desingularization'') by a family of $Spin(7)$ holonomy metrics on $X_8^{(B)} \cong \R^3 \times S^5$. This family of Ricci-flat metrics, conjectured in \cite{Gukov:2002zg} and only recently constructed in \cite{Foscolo:2019ilu}, can also be obtained as the total space of a circle bundle over the same $X_7$, but with a different locus of singular $S^1$ fibers
\be
\text{Phase B:} \qquad
M_4 \; = \; S^3 \times \R
\label{M4B}
\ee
The topology changing transition between these two coassociative submanifolds --- or, between the corresponding $Spin(7)$ manifolds $X_8^{(A)}$ and $X_8^{(B)}$ --- passes through the singular geometry, for which $M_4$ is a cone on $S^3 \sqcup S^3$.

\begin{figure}[ht]
	\centering
	\includegraphics[width=4.5in]{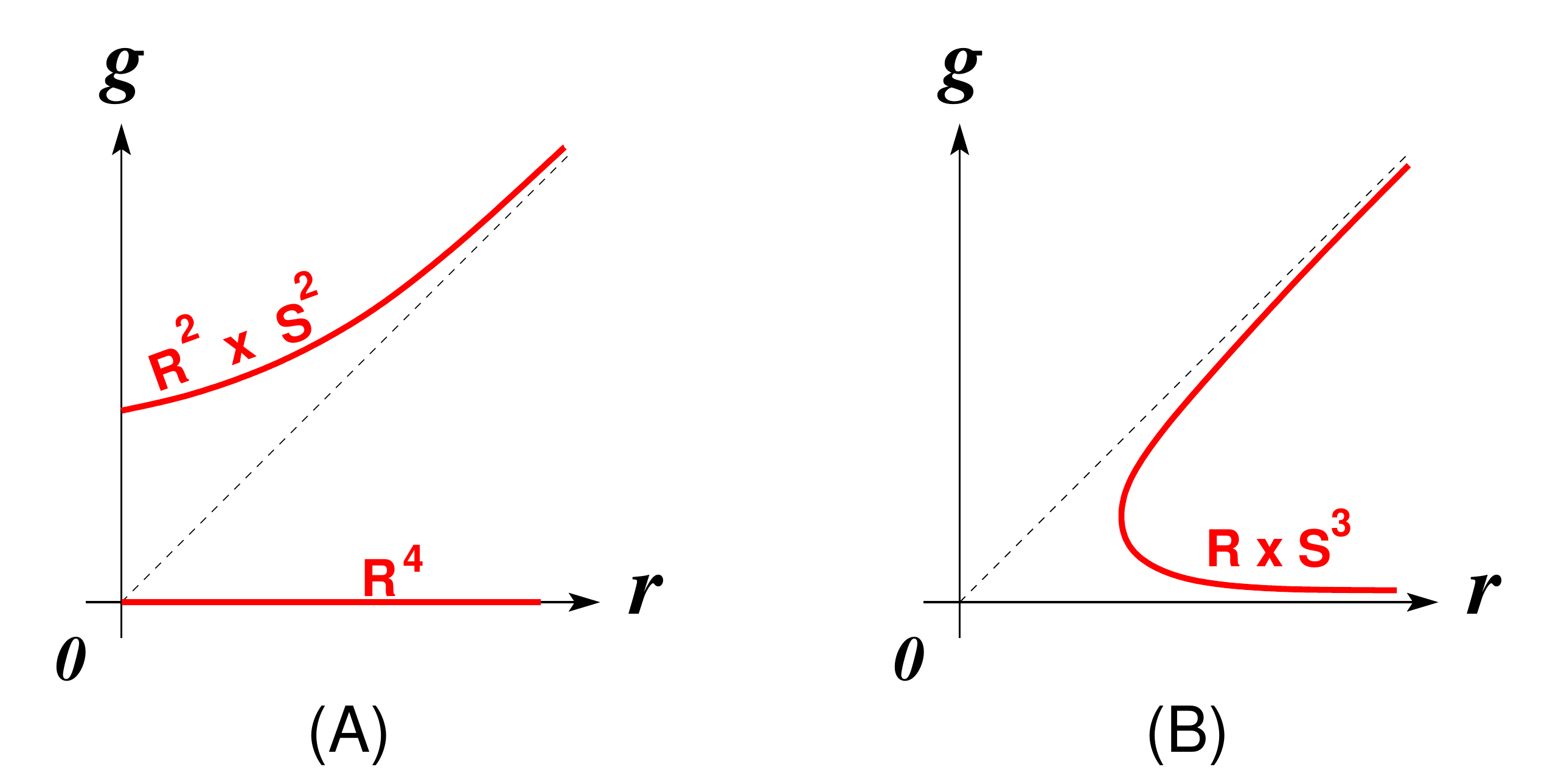}
	\caption{Two geometric phases of the coassociative 4-manifold \eqref{M4A}--\eqref{M4B} in $\R^3 \times \text{Taub-NUT}$. To illustrate only the homotopy type, we replace Taub-NUT by $\R^4$ and Spin bundle over $\cp^1 \cong S^2$ by $\R^2 \times S^2$.}
	\label{fig:coass-trans}
\end{figure}

A natural question, then, is: What is the physics of this transition? From the M-theory viewpoint, it requires studying the effective 3d $\CN=1$ theory obtained by compactification on $X_8$. From the perspective of type IIA theory on $X_7$, it requires understanding world-volume theory of the D6-brane supported on $\R^3 \times M_4$. Either way, the result of this analysis is a simple 3d $\CN=1$ theory \cite{Gukov:2002zg}:
\begin{center}
$U(1)$ gauge theory with Chern-Simons coupling at level $k = \pm \frac{1}{2}$\\ and a $\CN=1$ matter multiplet of charge $+1$
\end{center}
such that what we call phase A and phase B correspond to Higgs and Coulomb branches of this effective theory. A notable feature of this low-energy dynamics is the half-integer Chern-Simons term $k = \pm \frac{1}{2}$ that on D6-brane world-volume originates from the flux quantization condition \cite{Freed:1999vc}:
\be
\int\limits_{\cp^1} \frac{F}{2\pi} \; = \; k \in \Z + \frac{1}{2}
\label{FreedWitten}
\ee
and the fact that $\CO (+1) \to \cp^1$ is not Spin. In particular, because of this, the parity symmetry of the three-dimensional theory is spontaneously broken on the Higgs branch, which therefore in turn splits into two components labeled by $k = + \frac{1}{2}$ and $k = - \frac{1}{2}$.

\begin{table}
	\begin{centering}
		\begin{tabular}{|c||c|c|c|c|c|}
			\hline
			~phase~ & ~$X_8$~ & ~$M_4$~ & ~3d phase~ & ~parity~ & ~$U(1)_J$~ \tabularnewline
			\hline
			\hline
			$\phantom{\int^{\int^\int}} \text{A} \phantom{\int_{\int}}$ & $\R^4 \times \cp^2$ & $\R^2 \times S^2 \; \sqcup \; \R^4$ & Higgs & $\text{\sffamily X}$ & $\checkmark$
			\tabularnewline
			\hline
			$\phantom{\int^{\int^\int}} \text{B} \phantom{\int_{\int}}$ & $\R^3 \times S^5$ & $S^3 \times \R$ & Coulomb & $\checkmark$ & $\text{\sffamily X}$
			\tabularnewline
			\hline
		\end{tabular}
		\par\end{centering}
	\caption{\label{tab:3dphases} Two geometric phases of $Spin(7)$ holonomy manifolds $X_8$ correspond to Higgs and Coulomb phases of 3d $\CN=1$ effective field theory.}
\end{table}

\subsection*{2d phase transitions for $n=1$}

Naively, one might expect that the effective 2d $\CN=(0,2)$ theory on the fivebrane world-volume in our setup \eqref{M5branes} is similar to the effective theory of a D6-brane compactified on the same coassociative 4-manifold.
In particular, for the two choices \eqref{M4A}--\eqref{M4B} of $M_4$, with $n=1$, one might expect two phases, which we still continue calling phase A and phase B.

However, looking into the analysis a little more closely one quickly runs into questions, which indicate that the physics of M5-branes can be rather different from the physics of D-branes, even when they wrap the same coassociative submanifolds. For example, one question about the fivebrane system \eqref{M5branes} is: What is the analogue of the Freed-Witten anomaly \eqref{FreedWitten}? And, does it lead to two vacua in phase \eqref{M4A}?

Since the world-volume theory of a single fivebrane has a Lagrangian description, one might hope to deduce the effective 2d $\CN=(0,2)$ theory simply by following the standard rules of the Kaluza-Klein reduction. However, by looking at the KK spectrum in \cite{Dedushenko:2017tdw}, it is not immediately clear what the answer should be for $M_4 = S^3 \times \R$. This manifold has no homology in degrees up to 3, whereas all fields of 6d $(0,2)$ theory on the fivebrane world-volume are represented by differential forms of degree less than 3, even after the partial topological twist along $M_4$. So, naively, the resulting 2d $\CN=(0,2)$ theory looks completely empty in phase \eqref{M4B}. Another issue is that $M_4 = S^3 \times \R$ is non-compact, whereas the Kaluza-Klein spectrum summarized in Table~1 of \cite{Dedushenko:2017tdw} assumes compactness of $M_4$. So, how shall we even think of this compactification from 6d to 2d on a non-compact~$M_4$?

Another puzzle has to do with the fact that, in M-theory compactification on $Spin(7)$ manifold $X_8^{(B)} \cong \R^3 \times S^5$ a key role is played by the charged particle coming from M5-brane wrapped on the 5-sphere $S^5$. In the corresponding type IIA setup with a D6-brane supported on a coassociative 4-manifold $M_4 = S^3 \times \R$, this charged particle comes from a D4-brane with world-volume $\R \times D^4$, such that $\partial D^4 = S^3$. But, what is the corresponding analogue of this charged state in a setup with M5-branes on the same coassociative manifold $M_4 = S^3 \times \R$? The only candidate could be an M2-brane ending on the M5-brane, but the dimension of M2-brane world-volume is too small for this. On the other hand, in a different phase, where $M_4 \cong \R^2 \times S^2$, M2-branes supported on $D^3$ and ending on M5-brane along the $S^2 = \partial D^3$ produce instantons (local operators) in 2d $(0,2)$ theory. What is their~role?

Luckily, all these questions conveniently resolve one another. For example, in the case of 3d $\CN=1$ phases, the spontaneous symmetry breaking in phase A that leads to two vacua with $k = + \frac{1}{2}$ and $k = - \frac{1}{2}$ is accompanied by the existence of a domain wall that interpolates between these vacua. In M-theory on $\R^3 \times X_8^{(A)}$, this half-BPS domain wall is a M5-brane supported on $\R^2 \times \cp^2$, such that $\R^2 \subset \R^3$ and $\cp^2 \subset X_8^{(A)}$ is a topologically non-trivial calibrated (Cayley) 4-cycle. In the corresponding type IIA setup with a D6-brane supported on $\R^3 \times M_4$, this half-BPS domain wall is a D4-brane with world-volume $\R^2 \times D^3$, such that $\R^2 \subset D^3$ as before, and $\partial D^3 = S^2 \subset M^4$ is a topologically non-trivial 2-cycle.

Since, upon reduction on $S^1$, the 2-form field on fivebrane world-volume gives rise to a gauge field 1-form on D4-brane world-volume, one might expect that anomalous quantization condition \eqref{FreedWitten} on D4-brane world-volume comes from a similar anomaly for the 2-form. Moreover, one would also expect that, in case of M5-branes, this anomaly is responsible for spontaneous breaking of 1-form symmetry, since after compactification on $S^1$ it becomes the origin of a spontaneous breaking of ordinary symmetry on D4-brane theory. Both of these expectations are, in fact, correct, {\it cf.} \cite{Witten:1999vg,Gadde:2013sca}.

Indeed, just like ordinary symmetry breaking is accompanied by the existence of domain walls that interpolate between vacua (in our case, labeled by $k = + \frac{1}{2}$ and $k = - \frac{1}{2}$), 1-form symmetry breaking is accompanied by codimension-2 vortices. In our fivebrane setup \eqref{M5branes}, these half-BPS vortices are precisely M2-branes supported on $\{ \text{pt} \} \times D^3$ that were mentioned in one of the above questions. Note, that these M2-brane vortices only exist in phase \eqref{M4A}, same geometric phase that hosts half-BPS domain walls in 3d $\CN=1$ theory and has $w_2 (M_4) \ne 0$.

\begin{figure}[ht]
	\centering
	\includegraphics[width=3.7in]{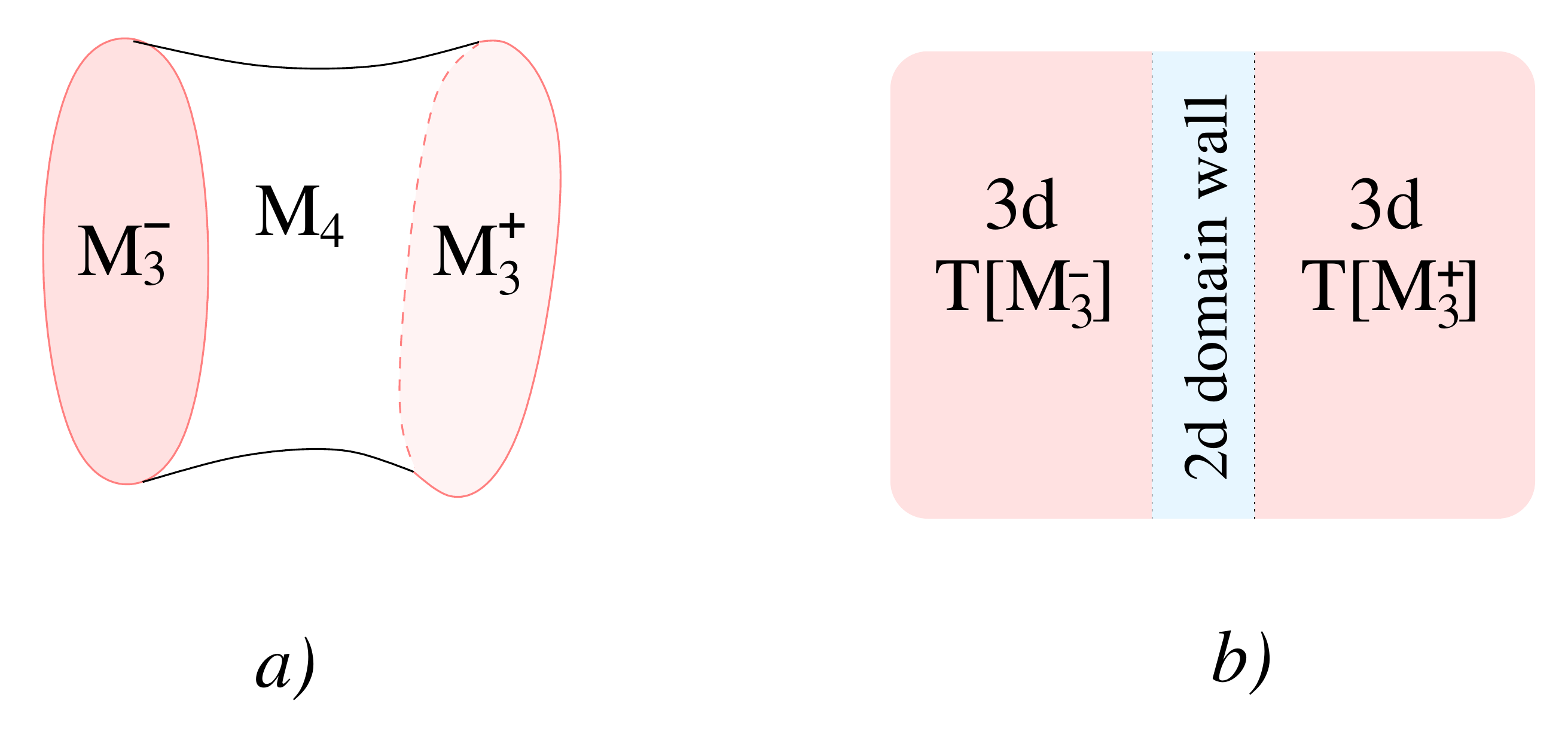}
	\caption{$(a)$ A cobordism between 3-manifolds $M_3^-$ and $M_3^+$ corresponds to $(b)$
		a 2d $\CN=(0,2)$ theory $T[M_4]$ on the domain wall (interface) coupled
		to 3d $\CN=2$ theories $T[M_3^-]$ and $T[M_3^+]$ on both sides \cite{Gadde:2013sca}.}
	\label{fig:M4d2dwall}
\end{figure}

Now let us address non-compactness of $M_4$, which also quickly turns on its head from an ugly problem into a nice feature. Indeed, since both choices of $M_4$ in phases \eqref{M4A} and \eqref{M4B} are non-compact, with two asymptotic regions $\approx \R \times S^3$, we should really think of these 4-manifolds as cobordisms from $M_3^- \cong S^3$ to $M_3^+ \cong S^3$. A ``compactification'' of a fivebrane on such cobordisms has a simple interpretation as an interface\footnote{equivalently, a non-dynamical domain wall} between 3d $\CN=2$ theories $T[M_3^-]$ and $T[M_3^+]$, illustrated in Figure~\ref{fig:M4d2dwall}.

Since $M_3^+ \cong M_3^-$, in our present problem we have the same theory $T[M_3^{\pm}]$ on both sides of the interface. In fact, we already encountered this 3d $\CN=2$ theory for general value of $n$ in our previous discussion, {\it cf.} \eqref{TLens}. Here we need its version with $n=1$ and $G=U(1)$, that is a $U(1)$ super-Chern-Simons theory at level 1 with an additional 3d $\CN=2$ free chiral multiplet. Therefore, it remains to identify two 2d $\CN=(0,2)$ interfaces in this theory, which correspond to \eqref{M4A} and \eqref{M4B}, respectively (or, more precisely, two phases of the {\it same} interface related by a 2d phase transition).

In one of these cases, the answer is simple: namely, in phase \eqref{M4B} the interface is ``trivial'' or ``fully transparent.'' Concretely, this means that it identifies the $U(1)_+$ and $U(1)_-$ gauge multiplets of theories $T[M_3^{+}]$ and $T[M_3^{-}]$, as well as free chiral multiplets on both sides. Note, in the 2d $\CN=(0,2)$ theory on the interface, $U(1)_+$ and $U(1)_-$ appear as global symmetries.

The phase A of our 2d interface is more interesting. (Recall, that the bulk 3d theory does not change, and it is only the 2d interface that undergoes a phase transition.) In this phase, our 2d interface is geometrically ``engineered'' by a compactification of 6d fivebrane theory on a 4-manifold with two connected componets,
\be
M_4 \; = \; M_4^- \, \sqcup \, M_4^+
\ee
where, according to \eqref{M4A},
\be
M_4^- \; = \;
\begin{matrix}
	\CO(+n) \\
	\downarrow \\
	\cp^1
\end{matrix}
\qquad , \qquad
M_4^+ \; = \; \text{TN}_{n}
\label{M4pmn}
\ee
written here, for convenience,\footnote{For the arguments to follow, it will be important to pay careful attention to orientations, which is easier if we restore general $n$ for a moment. As explained in \cite{Gadde:2013sca}, the manifolds $A_{n-1}$ and $\CO(-n) \to \cp^1$ have oppositely oriented boundaries and unique Spin structures. Therefore, they can be glued to form a closed 4-manifold which, however, is not Spin. In our case, $M_4^+$ and $M_4^-$ should be oriented in a way that allows a topology changing transition to $M_4 = \R \times L(n,1)$. This means if $M_4^+ = A_{n-1}$, then $M_4^-$ must be $\CO(+n) \to \cp^1$.} with general value of $n$.
Each of these components has one asymptotic region of the form $\R \times S^3$, which corresponds to either left or right side of Figure~\ref{fig:M4d2dwall}:
\be
\partial M_4^{\pm} \; = \; M_3^{\pm} \quad \left( \, \cong S^3 \right)
\ee
In other words, as our notations suggest, $T[M_4^+]$ couples to $T[M_3^+]$ and $T[M_4^-]$ couples to $T[M_3^-]$. Moreover, because the components of $M_4 = M_4^- \sqcup M_4^+$ are disjoint --- in fact, separated by a distance controlled by the parameter $\epsilon \sim \text{Vol} \left( S^2 \right)$, {\it cf.} \eqref{NS5-locus} --- the two ``halves'' of the interface, $T[M_4^+]$ and $T[M_4^-]$, become weakly interacting (decoupled) as~$\epsilon \to \infty$.

To summarize so far, our 2d interface $\CI$ has two branches, geometrically engineered by topological reduction on \eqref{M4A} and \eqref{M4B}. These branches are parametrized by $\text{Vol} (S^2)$ and $\text{Vol} (S^3)$, respectively. And, asymptotically, in the large volume limits, the interface becomes either fully transmissive or totally reflective, {\it cf.} \cite{Gadde:2013sca}:
\be
\begin{array}{lll}
\text{Phase A:}~~~ & \text{Maximally reflective}~\CI = T[M_4^+] \otimes T[M_4^-]~~~ & \text{as}~\text{Vol} (S^2) \to \infty \\
\text{Phase B:} & \text{Identity interface}~\CI = {\bf 1} & \text{as}~\text{Vol} (S^3) \to \infty
\end{array}
\label{2dMlimits}
\ee
Moreover, since for our choices of $M_4^{\pm}$ the individual 2d $\CN=(0,2)$ boundary theories $T[M_4^{\pm}]$ are known, it only remains to couple them in a consistent way, so that the combined 2d theory has two branches with the desired asymptotic behavior \eqref{2dMlimits}.

In phase A, the free chiral multiplet of 3d $\CN=2$ theory has Neumann boundary conditions on both sides of the interface, in $T[M_4^+]$ as well as in $T[M_4^-]$, meaning that in a decomposition to 2d $\CN=(0,2)$ chiral and Fermi multiplets the Fermi gets Dirichlet boundary conditions while the chiral obeys Neumann boundary conditions. A simple way to see this is to note that, in our general setup \eqref{M5branes}, each connected components of the M5-brane can move independently in the directions of $\R^4$ that are orthogonal to $\R^2$. For concreteness, we choose these directions to be $x^3$ and $x^4$.

In our present example \eqref{M4A}, the fivebrane has two connected components, $M_4^+$ and $M_4^-$, and therefore the corresponding 2d effective theories $T[M_4^+]$ and $T[M_4^-]$ each come equipped with a ``center of mass'' $(0,2)$ chiral multiplet \cite{Dedushenko:2017tdw}, whose (complex) scalar component parametrizes the motion along $x^3 + i x^4$. Naturally, we denote these two $(0,2)$ chiral multiplets in $T[M_4^+]$ and $T[M_4^-]$, respectively, by $\Phi_+$ and $\Phi_-$. These are precisely the restrictions (boundary values) of 3d $\CN=2$ chiral multiplet $\Phi_{3d}$ to the left and right boundaries of the 2d $\CN=(0,2)$ interface:
\be
\Phi_+ \; = \; \Phi_{3d} \big\vert_{\partial_+}
\qquad , \qquad
\Phi_- \; = \; \Phi_{3d} \big\vert_{\partial_-}
\label{2d3dchirals}
\ee
Note, in phase A, these two $(0,2)$ chiral multiplets are independent, whereas in phase B they are identified, {\it cf.} \eqref{2dMlimits}:
\be
\begin{array}{ll}
	\text{Phase A:}~~~ & \Phi_+ - \Phi_- \; = \; \epsilon \; = \; \text{Vol} (S^2) \cdot \exp \int\limits_{S^2} B \\
	\text{Phase B:} & \Phi_+ \; = \; \Phi_-
\end{array}
\label{S2scalars}
\ee
Here, through a minor abuse of notations, in phase A we also identified the difference of complex scalars in $\Phi_+$ and $\Phi_-$ with the geometric parameters of the compactification on \eqref{M4A}. In particular, $B$ denotes the 2-form field\footnote{whose field strength is self-dual} in 6d $(0,2)$ fivebrane theory

Note, the scalar fields in $T[M_4^+]$ and $T[M_4^-]$ parametrizing the motion in directions transverse to $\R^2 \subset \R^4$ can only be components of 2d $(0,2)$ chiral multiplets, which gives an {\it a priori} reason for \eqref{2d3dchirals} and for why the volume of $S^2 \subset M_4^-$ is complexified in \eqref{S2scalars}. The rotation in these directions, that is directions we chose to call $(x^3,x^4)$, is precisely the R-symmetry of 2d $\CN=(0,2)$ boundary supersymmetry algebra on the interface:
\be
U(1)_R \; = \; U(1)_{34}
\ee

In fact, for our $M_4^+$ and $M_4^-$ as in \eqref{M4pmn}, we know the complete description of the boundary theories $T[M_4^+]$ and $T[M_4^-]$, not just the sectors involving $(0,2)$ chiral fields $\Phi_{\pm}$. One of them, namely $T[M_4^+]$, we already encountered in section~\ref{sec:anomaly} when we introduced D6-branes in \eqref{D4NS5D6}. In fact, this is precisely how $M_4^+$ in our example, given by \eqref{generalM4M3}, ended up being the Taub-NUT space. And, the reason for incorporating D6-branes and modifying \eqref{M4viaSL} into \eqref{generalM4M3} was to introduce chiral fermions charged under the gauge symmetry of 3d $\CN=2$ theory so as to cancel the anomaly.
Therefore, according to that discussion, the 2d $\CN=(0,2)$ theory $T[M_4^+]$ in our example is the simplest theory of a D4-D6 brane intersection. In general, such intersection of $N$ D4-branes with $n$ D6-branes supports $nN$ Dirac fermions --- {\it i.e.} $nN$ Fermi multiplets in 2d $\CN=(0,2)$ language --- that transform in a bifundamental representation of $U(n) \times U(N)$, {\it cf.} \cite{Dijkgraaf:2007sw}.

From the viewpoint of 3d $\CN=2$ theory \eqref{TLens}, this $T[M_4^+]$ imposes a Dirichlet boundary condition on the $G=U(N)$ gauge field, whose ``edge modes'' (free Dirac fermions) realize $SU(n)$ current algebra at level $N$. This can be also seen directly from the physics of the 6d fivebrane theory on $M_4^+ = \text{TN}_n$, since $H^2 (\text{TN}_n, \Z)$ is the root lattice of $A_{n-1} = SU(n)$ and supports $n$ harmonic $L^2$-normalizable 2-forms. These 2-forms are all anti-self-dual, which means \cite{Dedushenko:2017tdw} that the corresponding Kaluza-Klein modes belong to the left-moving (non-supersymmetric) sector of the 2d $\CN=(0,2)$ theory $T[M_4^+]$. For our application, we are interested in the simplest case of this scenario, when $n=N=1$, so that the edge modes consist of only one Dirac fermion, a 2d $\CN=(0,2)$ Fermi multiplet that we call $\Psi_+$.

Similarly, $T[M_4^-]$ with $M_4^-$ as in \eqref{M4pmn} is also fairly well understood; it plays an important role in gluing operations on 4-manifolds and the corresponding chiral algebras. Since $M_4^-$ has positive-definite intersection form, we expect that edge modes should be represented by charged $(0,2)$ chiral multiplets, instead of Fermi multiplets that one has in a negative-definite case. This is indeed the case \cite{Dedushenko:2017tdw}, and from the direct Kaluza-Klein reduction in our example we find that $T[M_4^-]$ carries a $(0,2)$ chiral multiplet $P_-$ charged under the gauge symmetry $U(1)_-$ of the 3d $\CN=2$ theory.\footnote{This switch from Fermi to chiral multiplets, and vice versa, is very typical under a parity reversal in 3d $\CN=2$ theory~\cite{Gadde:2013sca} or under orientation reversal on the 4-manifold~\cite{Feigin:2018bkf}.} In the non-abelian case, that is for $N>1$, the boundary theory $T[M_4^-]$ is more interesting and carries the left-moving algebra called $\bar \CU$, see \cite{Feigin:2018bkf} for details.

Note, in the abelian case (that is, for $N=1$), the bulk 3d theory $T[M_3^+] \cong T[M_3^-]$ has no fermions charged under the corresponding gauge symmetry $U(1)_+$ or $U(1)_-$. Therefore, the only contribution to the 2d chiral anomaly from 3d bulk comes from the Chern-Simons term, which contributes $\pm n$ to the 't Hooft anomaly of $T[M_4^{\pm}]$. The sign of the anomaly is opposite for the two sides of the interface since reversing the orientation is equivalent to reversing the sign of the Chern-Simons term. These two contributions $+1$ and $-1$ from the 3d Chern-Simons coupling are neatly canceled by the anomalies of 2d $(0,2)$ multiplets $\Psi_+$ and $P_-$ charged under $U(1)_+$ and $U(1)_-$, respectively. In particular, the net anomaly is zero, which is crucial for the interface to become a trivial (identity) interface in phase B, where $U(1)_+ = U(1)_-$ becomes the gauge symmetry of the original 3d $\CN=2$ theory~\eqref{TLens}.

The last ingredient we need is a field whose scalar component can be identified with the volume of $S^3$ in phase \eqref{M4B}. Since 2d $\CN=(0,2)$ requires all scalars come in complex pairs, we propose that $\text{Vol} (S^3)$ is a (real) part of a 2d $\CN=(2,2)$ chiral multiplet, that is a pair of $(0,2)$ chiral and Fermi $(\Phi_0, \Psi_0)$:
\be
\Phi_0 \; = \; \text{Vol} (S^3) \cdot e^{i \theta}
\ee

Now we finally accounted for all key elements of the sought after 2d $\CN=(0,2)$ theory with phases~\eqref{2dMlimits}. Namely, it has two coupled sectors, $T[M_4^+]$ and $T[M_4^-]$, that correspond to the two sides of the interface in Figure~\ref{fig:M4d2dwall}. The field content of $T[M_4^+]$ includes a Fermi multiplet $\Psi_+$ and a $(0,2)$ chiral multiplet $\Phi_+$, whereas that of $T[M_4^-]$ contains two $(0,2)$ chiral multiplets $\Phi_-$ and $P_-$:
\be
\begin{array}{l@{\;}|@{\;}cccccc}
	& \Phi_+ & \Psi_+ & \Phi_- & P_- & \Phi_0 & \Psi_0 \\\hline
	U(1)_+ & 0 & +1 & 0 & 0 & 0 & 0 \\
	U(1)_- & 0 & 0 & 0 & -1 & 0 & 0
\end{array}
\label{Ifields}
\ee
There is an obvious candidate for the supersymmetric interaction of these fields, namely a $J$-term superpotential, that leads to the desired vacuum structure~\eqref{2dMlimits}:
\be
\int d \theta^+ \, \Phi_0 P_- \Psi_+ \, \big\vert_{\bar \theta^+ = 0}
\; + \;
\int d \theta^+ \, \left( \Phi_+ - \Phi_- \right) \Phi_0 \Psi_0 \, \big\vert_{\bar \theta^+ = 0}
\; + \; c.c.
\label{JforI}
\ee
In phase B, when a complex scalar in $\Phi_0$ gets a vev, this superpotential acts as a mass term for $P_-$ and $\Psi_+$, effectively setting these fields to zero. The variation with respect to $\Psi_0$ implies
\be
\left( \Phi_+ - \Phi_- \right) \Phi_0  \; = \; 0
\ee
so that either $\Phi_+ - \Phi_- \ne 0$ with $\Phi_0 = 0$ (phase A) or vice versa (phase B).

We believe the two-dimensional phase transition described here is in the same universality class as the phase transition in a simple type IIB brane model that we describe next. It would be nice to understand the relation between these brane models better, {\it e.g.} find a precise duality that relates the two.


\subsection*{Hanany-Witten type brane models}

Here we propose another brane system that realizes the field theory setup of the 3d $\CN=2$ gauge theory \eqref{TLens} with a half-BPS interface \eqref{Ifields}, thus making a full circle back to type IIB string theory and closer to the brane brick models \cite{Franco:2015tya,Franco:2016nwv,Franco:2016qxh} that were our original inspiration.

\begin{figure}[ht]
	\centering
	\includegraphics[width=4.5in]{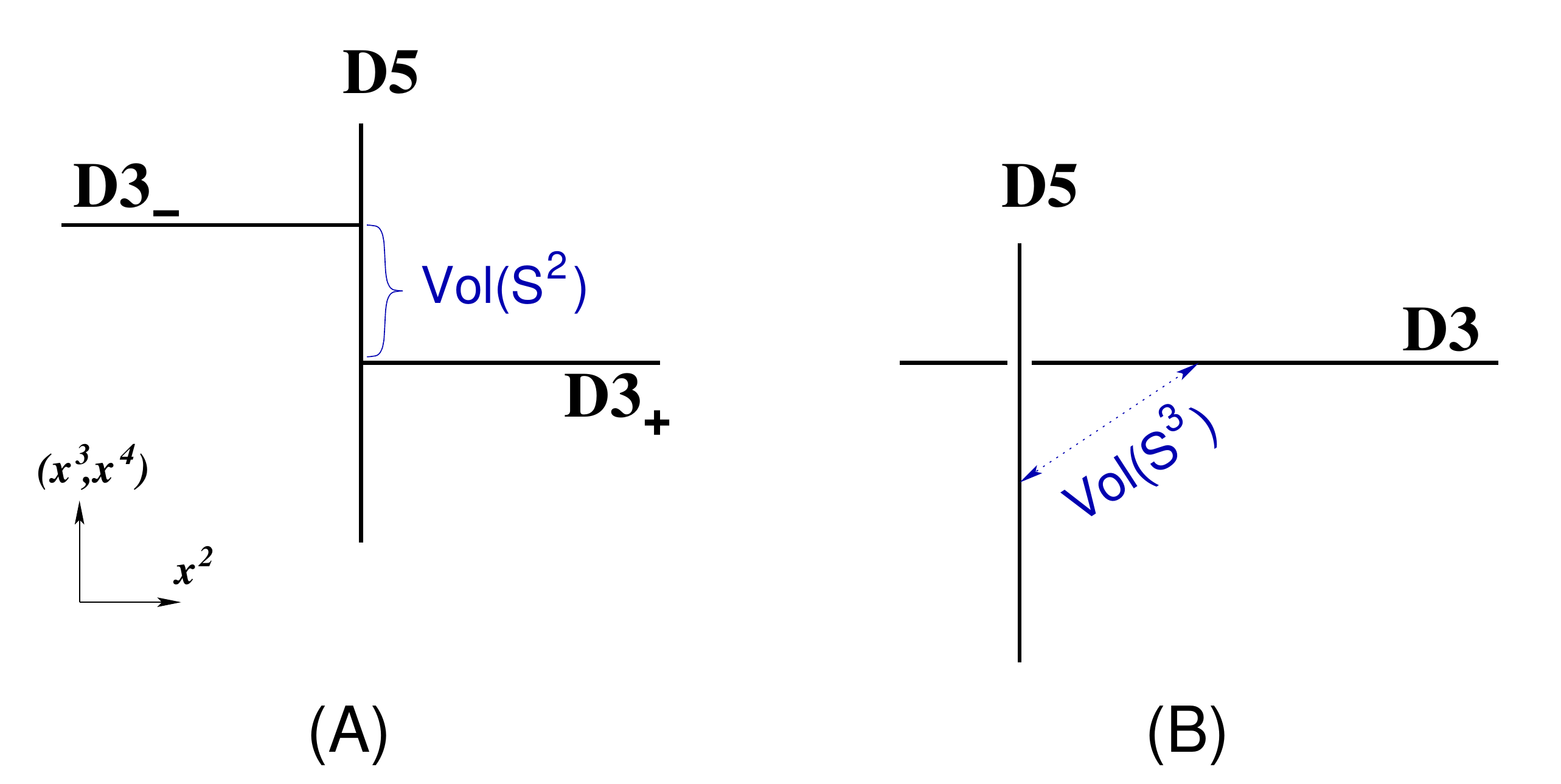}
	\caption{Two phases of a 2d $\CN=(0,2)$ interface \eqref{2dMlimits} in a 3d $\CN=2$ theory \eqref{TLens}.}
	\label{fig:IIBbranes}
\end{figure}

Following \cite{Hanany:1996ie}, we can consider a stack of $N$ D3-branes suspended between two fivebranes in type IIB string theory. We choose one of the fivebranes to be a NS5-brane, with world-volume along directions 012345, and the second fivebrane to be of type $(p,q)$ and oriented along directions 01234$[59]_{\theta}$, where $[59]_{\theta}$ indicates a rotation in $(x^5,x^9)$ plane by angle $\theta$. It is well known \cite{Ohta:1997fr,Kitao:1998mf,Bergman:1999na} that such brane configuration carries 3d $\CN=2$ gauge theory \eqref{TLens} on the D3-brane world-volume, provided that
\be
n \; = \; \frac{p}{q} \; = \; \tan \theta
\ee
is an integer (equal to the value of the Chern-Simons coefficient). We assume the world-volume of D3-branes to be along the directions $(x^0,x^1,x^2)$ and $x^6$:
\begin{align}\notag
\hbox{NS5} &: \quad 012345 \cr
\hbox{$(p,q)$5} &: \quad 01234[59]_{\theta} \cr
\hbox{D3}_+ &: \quad 012_+~~~~6 \cr
\hbox{D3}_-  &: \quad 012_-~~~~6 \cr
\hbox{D5}  &: \quad 01~~~3456
\end{align}
where in the last line we also included a D5-brane localized along the $x^2$ direction. This D5-brane breaks half of the supersymmetries of the 3d $\CN=2$ theory on D3-brane, and therefore creates a codimension-1 half-BPS defect.\footnote{A similar realization of half-BPS defects in 3d $\CN=2$ super-Chern-Simons theory on D3-branes was considered in \cite{Armoni:2015jsa}, though details of the corresponding brane models appear to be different.} Also indicated in this Table, and illustrated in Figure~\ref{fig:IIBbranes}, is the possibility of D3-brane(s) splitting into two pieces along the D5-brane.

Since the only directions shared by all three fivebranes in our IIB system are $(x^0,x^1)$ and $(x^3,x^4)$, the D3-branes can only separate along $(x^3,x^4)$,
\be
\begin{array}{ll}
	\text{Phase A:}~~~ & \Delta x^3 + i \Delta x^4 \; = \; \epsilon \; = \; \Phi_+ - \Phi_- \\
	\text{Phase B:} & \Delta x^7 + i \Delta x^8 \; = \; \Phi_0
\end{array}
\ee
Here we identified this separation with the parameters in~\eqref{S2scalars} since the physics on the branch parametrized by $\epsilon \ne 0$ matches that of phase A in the above discussion.
Similarly, phase B of the 2d $\CN=(0,2)$ theory at the D3-D5 intersection corresponds to moving the D5-brane along the $(x^7,x^8)$ directions, as illustrated on the second panel of Figure~\ref{fig:IIBbranes}.

It should not be too difficult to generalize this analysis of brane configurations --- in M-theory as well as in type IIB string theory --- to arbitrary values of $n$ and $N$. More generally, it is natural to ask: What kinds of topology changing transitions can coassociative 4-manifolds have?
We leave these interesting problems to future work.


\acknowledgments{It is pleasure to thank Robert Bryant, Lorenzo Foscolo, Sheldon Katz, Rafe Mazzeo, Jeffrey Meier, Grigory Mikhalkin, and Nathan Seiberg for useful discussions.
The work of S.F. is supported by the National Science Foundation grant PHY-1820721.
The work of S.G. is supported by the U.S. Department of Energy, Office of Science, Office of High Energy Physics, under Award No. DE-SC0011632, and by the National Science Foundation under Grant No. NSF DMS 1664240.
The work of S.L. is supported by Samsung Science and Technology Foundation under Project Number SSTF-BA1402-08.
}


\appendix

\section{Supersymmetry conditions for D6-branes}
\label{sec:D6}

Here we show that brane configuration \eqref{D4NS5} in type IIA string theory allows adding
D6-branes supported on special Lagrangian submanifolds in $X$, without breaking supersymmetry further.
Moreover, and especially important for our applications, the orientation of extra D6-branes
needs to be strictly correlated with the orientation of D4-branes; namely, if D4-branes
are calibrated by $\text{Re} (e^{i \theta} \Omega)$, then D6-branes must be calibrated by $\text{Im} (e^{i \theta} \Omega)$.

For purposes of analyzing supersymmetry conditions, we can replace $X$ by $\C^3$, with its flat Calabi-Yau structure \eqref{CYflat}, and imitate $S$ by a triple of complex hyperplanes supported at $z_1 = 0$, $z_2=0$, and $z_3 = 0$,
{\it cf.} \eqref{SPinC3}. In these conventions, D4-branes are calibrated by the 3-form $\text{Re} (\Omega)$, and the brane configuration looks as follows:
\begin{align}
\begin{array}{l|cc|cccccc|cc}
& 0 & 1 & 2 & 3 & 4 & 5 & 6 & 7 & 8 & 9
\\
\hline
\mbox{NS5} & \times & \times& \times & \times & \times & \times & & & & 
\\
\mbox{NS5$'$} & \times & \times& \times & \times & & & \times & \times & & 
\\
\mbox{NS5$''$} & \times & \times &  &  & \times & \times & \times & \times & & 
\\
\mbox{D4} & \times & \times & \times & & \times & & \times & & & 
\\
\end{array}
\label{brane-config}
\end{align}
The supersymmetry condition for a NS5-brane with world-volume along the directions $012345$ is given by $\e_{L,R} = - \G^{012345} \e_{L,R}$, and similarly for the other two branes NS5$'$ and $NS5''$. Here, $\e_L$ and $\e_R$ are 10d spinors of left and right chirality:
\begin{align}
\e_L = - \G^{0123456789} \e_L 
\,, 
\quad 
\e_R = +\G^{0123456789} \e_R \,.
\end{align}
Combining the supersymmetry condition for a D4-brane, $\e_L = \G^{01246} \e_R$, with the above mentioned condition for a NS5-brane, after simple gamma-matrix algebra we obtain $\e_L = \G^{0124789} \e_R$. This is precisely the supersymmetry condition for a D6-brane with world-volume along the directions $0124789$. Similarly, combining the D4-brane supersymmetry condition with those for NS5$'$ and NS5$''$ branes, we learn that any of the following D6-branes can be introduced into the brane configuration~\eqref{brane-config} without breaking supersymmetry further \cite{GarciaCompean:1998kh}:
\begin{align}
\begin{array}{l|cc|cccccc|cc}
& 0 & 1 & 2 & 3 & 4 & 5 & 6 & 7 & 8 & 9
\\
\hline
\mbox{D6} & \times & \times & \times & & \times & & & \times & \times & \times
\\
\mbox{D6$'$} & \times & \times & \times & & & \times & \times & & \times & \times
\\
\mbox{D6$''$} & \times & \times & & \times & \times & & \times & & \times & \times
\end{array}
\end{align}
Note, all of these D6-branes meet the original D4-branes along two directions inside $X$, precisely as special Lagrangian submanifolds calibrated by the 3-form $\text{Im} (\Omega)$ (with our convention that D4-branes are calibrated by $\text{Im} (\Omega)$). The other set of D6-branes, with world-volume along directions $0135789$ is also calibrated by $\text{Im} (\Omega)$ and, as can be seen directly via gamma-matrix algebra, does not break supersymmetry further.

\section{Calibration condition}

Here we show how to derive the ODE \eqref{ode-final} from the calibration condition $\Phi|_{M_4}=0$, $\Psi|_{M_4} = d(\mbox{vol})_{M_4}$ and the $SO(3)$-invariant ansatz introduced in section \ref{sec:ODE}. For concreteness, we focus on the case $n=2$, but 
generalization to other $n$ is straightforward.

\paragraph{Taub-NUT}

Introducing $n$ D6-branes replaces $\mathbb{C}^3 \times S^1$  by $\mathbb{R}^3 \times \text{TN}_n$, where TN$_n$ is an $n$-centered Taub-NUT.
Recall that the Taub-NUT metric takes the form 
\begin{align}
ds^2(\text{TN}_n) = H d\vec{x}^2 + H^{-1}(d\tilde{\psi} + \chi)^2 \,,
\quad d\chi =  *_3 dH\,, 
\label{metric-TN-copy}
\end{align}
where $*_3$ is the Hodge dual with respect to the flat metric of $\mathbb{R}^3(\vec{x}) = \text{TN}_n/S^1$. The angle variable $\tilde{\psi}$ has period $4\pi$. We choose the harmonic function $H$ 
to preserve the $SO(3)$ symmetry in $\mathbb{R}^3(\vec{x})$\, 
\begin{align}
H = B + \frac{n}{r} \,, 
\quad \chi =  n \cos\theta d\phi \,.
\end{align}
The overall orientation of $\mathbb{R}^3 \times \text{TN}_n$ is fixed by
\begin{align}
d(\text{vol})_{\mathbb{R}^3 \times \text{TN}_n} = dy^{123}  \wedge H dx^{123} \wedge (d\tilde{\psi} + \chi)   =  d(\text{vol})_{\mathbb{R}^3} \wedge d(\text{vol})_{\text{TN}_n}\,.
\label{orien-R3TNn}
\end{align}

When $r \gg 1$, the TN metric asymptotes to $\mathbb{R}^3 \times S^1$ 
with the circumference of $S^1$ approaching $4\pi$. When $r \ll 1$, the constant term in $H$ is suppressed and the metric approaches that of $\mathbb{C}^2/\mathbb{Z}_n$, 
\begin{align}
ds^2 \approx d\rho^2 + \frac{\rho^2}{4} \left[ d\theta^2 + \sin^2\theta d\phi^2 + (d\psi + \cos\theta d\phi)^2 \right]
\,, 
\quad \psi := \frac{1}{n} \tilde{\psi}\,, 
\quad
\rho^2 := (4n) r\,. 
\label{r-vs-rho}
\end{align}

\paragraph{Associative and co-associative}

To obtain the associative 3-form $\Phi$ for $\mathbb{R}^3(\vec{y}) \times \text{TN}_n(\vec{x},\tilde{\psi})$, we begin with \eqref{coass-cy} and make the replacement, 
\begin{align}
dx^i \rightarrow H^{\frac{1}{2}} dx^i \,,
\quad 
d\psi \rightarrow H^{-\frac{1}{2}} (d\tilde{\psi} + \chi) \,,
\end{align}
so that 
\begin{align}
\begin{split}
\Psi &=  H \left( \frac{1}{2} J \wedge J \right)
+ \left( H dx^{123}  - dy^{ij}\wedge dx^k \right) \wedge (d\tilde{\psi} + \chi) 
= d(\text{vol})_{\text{TN}_n} - dy^{ij} \wedge \omega^k \,,
\\
\Phi &= J \wedge (d\tilde{\psi} + \chi) + dy^{123} - H \left( dy^k \wedge dx^{ij} \right) = 
d(\text{vol})_{\mathbb{R}^3_y} - dy^k \wedge \omega^k \,.
\label{coass-tn}
\end{split}
\end{align}
We introduced the well-known self-dual 2-forms on TN$_n$, \begin{align}
\omega^k = H dx^{ij} +  dx^k \wedge (d\tilde{\psi} + \chi) \,, 
\quad 
\quad 
d\omega^k = 0 \,,
\quad 
*_\text{TN} (\omega^k) = + \omega^k \,,
\label{TN-omega-def}
\end{align}
and a few short-hand notations:
\begin{align}
\begin{split}
dx^{123} &= dx^1 \wedge dx^2 \wedge dx^3 \,,
\\
dx^k \wedge dy^{ij} &= dx^1 \wedge (dy^2 \wedge dy^3) + dx^2 \wedge (dy^3 \wedge dy^1) + dx^3 \wedge (dy^1 \wedge dy^2) \,,
\\
dy^k \wedge \omega^k &= dy^1 \wedge \omega^1 + dy^2 \wedge \omega^2 + dy^3 \wedge \omega^3 \,.
\end{split}
\end{align}
Other notations are understood in similar ways. 
Clearly, $d\omega^k=0$ implies 
$d\Psi = 0 = d\Phi$. 

\paragraph{Embedding ansatz}

We embed $M_4$ into $\mathbb{R}^3 \times \text{TN}_2$ by expressing $\vec{y}$ as functions of $(\vec{x},\tilde{\psi})$. It is more convenient to use polar coordinates $(r,\theta,\phi,\psi)$, where $r = |\vec{x}|$ 
and $(\theta,\phi,\psi)$ are the standard angle coordinates on $S^3$. The $\psi$ here is related to $\tilde{\psi}$ in \eqref{metric-TN-copy} by 
\begin{align}
\tilde{\psi} = n \psi  = 2\psi \,.
\end{align}

We choose specific coordinates in the unit quaternion/spinor description of $S^3$, 
\begin{align}
q = 
\begin{pmatrix}
u_1 & -\bar{u}_2 \\
u_2 & \bar{u}_1 
\end{pmatrix}\,, 
\quad 
u = 
\begin{pmatrix}
u_1  \\
u_2
\end{pmatrix} = 
\begin{pmatrix}
\cos(\theta/2)e^{i(-\psi-\phi)/2}  \\
\sin(\theta/2)e^{i(-\psi+\phi)/2}
\end{pmatrix} \,, 
\quad 
\tilde{u} = 
\begin{pmatrix}
-\bar{u}_2 \\
\bar{u}_1 
\end{pmatrix}\,.
\end{align}
The associated invariant 1-forms are
\begin{align}
\begin{split}
\vec{\tau} &= i \,\Tr\left( \vec{\sigma} q^\dagger dq \right) 
\\
&= (\sin\psi d\theta - \cos\psi \sin\theta d\phi , \cos\psi d\theta + \sin\psi \sin\theta d\phi , d\psi + \cos\theta d\phi) \,.
\end{split}
\end{align}
They satisfy
\begin{align}
d\tau_i = - \frac{1}{2} \epsilon^{ijk} (\tau_j \wedge \tau_k) \,, 
\quad 
\tau_1 \wedge \tau_2 \wedge \tau_3 = \sin\theta (d\theta \wedge d\phi \wedge d\psi) \,.
\end{align}

The $\vec{x}$ coordinates are mapped to $(r,\theta,\phi)$ coordinates in the usual way, 
\begin{align}
\vec{x} = r \,\hat{m} \,,
\quad 
\hat{m} = u^\dagger \vec{\sigma} u = (\sin\theta \cos\phi, \sin\theta \sin\phi, \cos\theta) \,, 
\label{coord-x}
\end{align}
such that 
\begin{align}
d\vec{x}^2 = dr^2 + r^2 d\hat{m}^2 = dr^2 + r^2 (\tau_1^2 +\tau_2^2) = dr^2 + r^2 (d\theta^2 + \sin^2\theta d\phi^2) \,.
\label{dx2-m}
\end{align}
As stated in \eqref{ansatz-y}, we look for an ansatz of the form
\begin{align} 
\vec{y} = g(r) \hat{n}(\theta, \phi,\psi) \,,
\quad
\hat{n} \cdot \hat{n} = 1\,,
\quad
\hat{m} \cdot \hat{n} = 0 \,.
\label{ansatz-y-copy}
\end{align}
Such a fibration is readily achieved by means of the spinor
\begin{align}
v = \frac{1}{\sqrt{2}} (u + \tilde{u}) \,,
\end{align}
such that
\begin{align}
\hat{n} = v^\dagger \vec{\sigma} v = 
\cos\psi ( \cos\theta  \cos\phi  , \cos\theta \sin\phi , -\sin\theta ) 
 + \sin\psi ( -\sin\phi , \cos\phi , 0 ) \,.
\end{align}
We note that, somewhat similarly to \eqref{dx2-m}, 
\begin{align}
d\hat{n}^2 = \tau_2^2 + \tau_3^2\,.
\end{align}

\paragraph{Calibration condition}

We want $M_4$ to be a coassociative 4-cycle, that is, 
\begin{align}
\Phi|_{M_4} = 0 \,, 
\quad 
\Psi|_{M_4} = d(\text{vol})_{M_4} \,.
\end{align}
Our goal is to obtain a first-order non-linear ODE for $g(r)$ from the first condition.

The angular part is fixed by the $SO(3)\times U(1)$ symmetry. To see it clearly, it is useful to rewrite $dm^k$ and $dn^k$as linear combinations of $\tau_i$. In a local orthonormal frame spanned by $\hat{m}$, $\hat{n}$ and $\hat{\ell} := \hat{m} \times \hat{n}$, we have
\begin{align}
dm^k = n^k \tau_2 - \ell^k \tau_1  \,,
\quad 
dn^k = \ell^k \tau_3-m^k \tau_2  \,, 
\quad 
d\ell^k = m^k \tau_1 - n^k \tau_3 \,.
\end{align}
It follows that 
\begin{align}
\begin{split}
m^i dm^j - m^j dm^i = n^k \tau_1 + \ell^k \tau_2 \,,
& \quad 
dm^i \wedge dm^j = m^k (\tau_1 \wedge \tau_2) \,,
\\
n^i dn^j - n^j dn^i = \ell^k \tau_2 + m^k \tau_3 \,,
& \quad 
dn^i \wedge dn^j = n^k (\tau_2 \wedge \tau_3) \,.
\end{split}
\end{align}

Let us compute the pull-back of $\Phi$ onto $M_4$: 
\begin{align}
\Phi &= dy^{123} - dy^k \wedge \omega^k 
= dy^{123} - dy^k \wedge \left( H dx^{ij} + 2 \, dx^k \wedge \tau_3\right)\,.
\end{align}
With the coordinates \eqref{coord-x} and the ansatz \eqref{ansatz-y-copy}, we have
\begin{align}
\begin{split}
& dx^k = m^k dr  + r (dm^k) = m^k dr + r (n^k \tau_2 - \ell^k \tau_1) \,,
\\
& dy^k = g' n^k dr  + g (dn^k)  =g' n^k dr  + g (\ell^k \tau_3 -m^k \tau_2 )  \,, 
\end{split}
\end{align}
so that `
\begin{align}
dx^{123} = r^2 dr (\tau_1 \wedge \tau_2) \,, \quad
dy^{123} =  g^2 g' dr (\tau_2 \wedge \tau_3)   \,. 
\end{align}
The two-forms $\omega_k$ get pulled back to 
\begin{align}
\begin{split}
\omega_k &= H r dr \wedge (m^i dm^j - m^j dm^i) +  H r^2 (dm^i  \wedge dm^j)   
\\
& \quad
+ 2 m^k dr \wedge \tau_3 + 2r \,dm^k \wedge \tau_3
\\
&= dr \wedge \left[ (Hr)(n^k \tau_1 +\ell^k \tau_2) 
+ 2m^k \tau_3 
 \right] 
\\
&\quad +
\left[ 
 (Hr^2) m^k (\tau_1 \wedge \tau_2) + (2r) \{ \ell^k (\tau_3 \wedge \tau_1) 
+n^k (\tau_2\wedge \tau_3 )
\}
\right]\,.
\end{split}
\label{omega-pullback}
\end{align}

In $\Phi$, one may naively expect a $(\tau_1 \wedge \tau_2 \wedge \tau_3)$ term 
and three $dr \wedge (\tau_i \wedge \tau_j)$ terms. It turns out that all but the $dr \wedge (\tau_2 \wedge \tau_3)$ terms vanish. Collecting all contributions, and demanding $\Phi|_{M_4} = 0$, we finally obtain the desired ODE for $n=2$:
\begin{align}
g^2 g' - (Hr)g - 2r g'  - 2g  = 0 
\quad 
\Longleftrightarrow 
\quad 
\boxed{ g' = \frac{(Hr+2)g}{g^2-2r} } \,.
\label{ode-final-copy}
\end{align}


\bibliographystyle{JHEP}
\bibliography{classH}

\end{document}